\documentclass[useAMS,usenatbib]{mn2e}

\usepackage[T1]{fontenc}
 \usepackage{pslatex}
\usepackage{amsmath}
\usepackage{amsfonts}
\usepackage{color}
\usepackage{url}
\usepackage{graphicx}
\usepackage{subfigure}
\usepackage{amssymb}
\usepackage{soul}
\usepackage{booktabs}
\usepackage{array}
\usepackage[utf8]{inputenc}
\inputencoding{latin1}
\inputencoding{utf8}

{\newif\ifnotend
\notendtrue
\def\veclist{ABCDEFGHIJKLMNOPQRSTUVWXYZabcdefghijklmnopqrstuvwxyz.}
\def\top#1#2.{#1}
\def\tail#1#2.{#2.}
\loop\expandafter\xdef\csname v\expandafter\top\veclist\endcsname%
{{\noexpand\bf\expandafter\top\veclist}}
\edef\veclist{\expandafter\tail\veclist}
\if\veclist.\notendfalse\fi\ifnotend\repeat}
\def\e{{\rm e}}

\def\E{{\cal E}}

\mathchardef\mhyphen="2D

\title[SSC radiation from turbulent plasmas]{Synchrotron-Self-Compton radiation from magnetically-dominated turbulent plasmas in relativistic jets}

\author[Sobacchi, Sironi and Beloborodov]{Emanuele Sobacchi$^{1}$\thanks{E-mail: es3808@columbia.edu}, Lorenzo Sironi$^1$ and Andrei M. Beloborodov$^{2,3}$\\
$^1$ Department of Astronomy and Columbia Astrophysics Laboratory, Columbia University, 550 West 120th Street New York, NY 10027, USA\\
$^2$ Physics Department and Columbia Astrophysics Laboratory, Columbia University, 538 West 120th Street New York, NY 10027, USA \\
$^3$ Max Planck Institute for Astrophysics, Karl-Schwarzschild-Str. 1, D-85741, Garching, Germany
}

\begin{document}

\date{}

\def\p{\partial}
\def\E{\textbf{E}}
\def\B{\textbf{B}}
\def\j{\textbf{j}}
\def\s{\textbf{s}}
\def\e{\textbf{e}}

\newcommand{\di}{\mathrm{d}}
\newcommand{\bfx}{\mathbf{x}}
\newcommand{\bfe}{\mathbf{e}}
\newcommand{\vlos}{\mathrm{v}_{\rm los}}
\newcommand{\Tspin}{T_{\rm s}}
\newcommand{\Tb}{T_{\rm b}}
\newcommand{\degree}{\ensuremath{^\circ}}
\newcommand{\Th}{T_{\rm h}}
\newcommand{\Tc}{T_{\rm c}}
\newcommand{\bfr}{\mathbf{r}}
\newcommand{\bfv}{\mathbf{v}}
\newcommand{\bfu}{\mathbf{u}}
\newcommand{\pc}{\,{\rm pc}}
\newcommand{\kpc}{\,{\rm kpc}}
\newcommand{\Myr}{\,{\rm Myr}}
\newcommand{\Gyr}{\,{\rm Gyr}}
\newcommand{\kms}{\,{\rm km\, s^{-1}}}
\newcommand{\de}[2]{\frac{\partial #1}{\partial {#2}}}
\newcommand{\cs}{c_{\rm s}}
\newcommand{\rb}{r_{\rm b}}
\newcommand{\rqu}{r_{\rm q}}
\newcommand{\bfOmega}{\pmb{\Omega}}
\newcommand{\bfOmegap}{\pmb{\Omega}_{\rm p}}
\newcommand{\bfXi}{\boldsymbol{\Xi}}

\maketitle

\begin{abstract}
Relativistic jets launched by rotating black holes are powerful emitters of non-thermal radiation. Extraction of the rotational energy via electromagnetic stresses produces magnetically-dominated jets, which may become turbulent. Studies of magnetically-dominated plasma turbulence from first principles show that most of the accelerated particles have small pitch angles, i.e. the particle velocity is nearly aligned with the local magnetic field. We examine synchrotron-self-Compton radiation from anisotropic particles in the fast cooling regime. The small pitch angles reduce the synchrotron cooling rate and promote the role of inverse Compton (IC) cooling, which can occur in two different regimes. In the Thomson regime, both synchrotron and IC components have soft spectra, $\nu F_\nu\propto\nu^{1/2}$. In the Klein-Nishina regime, synchrotron radiation has a hard spectrum, typically $\nu F_\nu\propto\nu$, over a broad range of frequencies. Our results have implications for the modelling of BL Lacs and Gamma-Ray Bursts (GRBs). BL Lacs produce soft synchrotron and IC spectra, as expected when Klein-Nishina effects are minor. The observed synchrotron and IC luminosities are typically comparable, which indicates a moderate anisotropy with pitch angles $\theta\gtrsim0.1$. Rare orphan gamma-ray flares may be produced when $\theta\ll0.1$. The hard spectra of GRBs may be consistent with synchrotron radiation when the emitting particles are IC cooling in the Klein-Nishina regime, as expected for pitch angles $\theta\sim0.1$. Blazar and GRB spectra can be explained by turbulent jets with a similar electron plasma magnetisation parameter, $\sigma_{\rm e}\sim10^4$, which for electron-proton plasmas corresponds to an overall magnetisation $\sigma=(m_{\rm e}/m_{\rm p})\sigma_{\rm e}\sim10$.
\end{abstract}

\begin{keywords}
gamma-ray bursts -- BL Lacertae objects: general -- radiation mechanism: non-thermal -- plasmas -- turbulence
\end{keywords}


\section{Introduction}

Relativistic jets from accreting black holes are powerful emitters of non-thermal radiation. Examples include Gamma-Ray Bursts (GRBs) \citep[e.g.][]{Piran2004, KumarZhang2015} and blazars \citep[e.g.][]{UrryPadovani1995, Blandford2019}.

Relativistic jets may be launched by a universal physical process, in which the rotational energy of the black hole is extracted through electromagnetic stresses \citep[e.g.][]{BlandfordZnajek1977, Komissarov2007, Tchekhovskoy2011}. This process produces magnetically-dominated jets, where the magnetic energy density exceeds the rest mass energy density of the plasma. Since there is a huge separation of scales between the transverse scale of the jet and the kinetic scales of the plasma, turbulence is a natural candidate to dissipate the magnetic energy and accelerate a population of non-thermal particles.

Since GRBs and blazars convert a similarly large fraction of the jet energy into gamma-rays \citep[e.g.][]{Nemmen2012}, it is natural to consider fast cooling conditions, i.e. the emitting particles radiate their energy on short timescales compared with the dynamical time of the jet expansion. When most of the jet energy is stored in the magnetic fields, synchrotron emission is usually expected to be the dominant cooling channel. Then fast cooling particles produce a soft synchrotron spectrum, $\nu F_\nu\propto \nu^\alpha$ with $\alpha=1/2$. For GRBs, this prediction of the synchrotron model is problematic, as the observed bursts show harder spectra with $\alpha\sim 1$ \citep[e.g.][]{Preece+2000, Kaneko+2006, Nava+2011, Gruber+2014}.

The hard GRB spectra generally favour photospheric emission models, where the peak of the spectrum is formed by multiple Compton scattering during the opaque stage of the jet expansion \citep[for a review, see e.g.][]{BeloborodovMeszaros2017}. Some GRBs appear to have a clear photospheric origin \citep[e.g.][]{Ryde+2010}. However, for many other GRBs the emission mechanism is not established. It is possible that in many GRB jets the dissipation occurs in the optically thin zone, and synchrotron dominates the observed emission \citep[e.g.][]{Oganesyan+2019, Burgess+2020}. Polarisation of the prompt radiation may help discriminate between different emission models \citep[e.g.][]{Lundman+2018, Gill+2020}, however observations using different instruments are not yet conclusive \citep[e.g.][]{Yonetoku+2011, Yonetoku+2012, Burgess+2019, Chand+2019, Chattopadhyay+2019, Sharma+2019, Zhang+2019, Kole+2020}.

The observed spectral slopes remain an important constraint for GRB and blazar models. For blazars, the emission is almost certainly due to synchrotron and inverse Compton \citep[e.g.][]{Maraschi1992, Sikora1994}. The spectrum is softer than for GRBs, and the typical slope, $\alpha\sim 1/2$, may be consistent with the standard fast cooling scenario. Although a common dissipation process in blazars and GRBs is an attractive possibility, one immediate challenge for such a model is to explain the spectral difference.\footnote{Several authors argued that magnetic energy dissipation in GRB jets provides a continuous source of heating, which may prevent particles from cooling down by radiative losses \citep[e.g.][]{ZhangYan2011, BeniaminiPiran2014, Beniamini+2018, Xu+2018}. The resulting synchrotron spectrum is harder than in the standard scenario where the heating/acceleration process is impulsive.} This issue is investigated in the present paper.

In recent years, increased computational capabilities made it possible to study non-thermal particle acceleration in magnetically-dominated turbulence from first principles \citep[e.g.][]{Zhdankin+2017, Zhdankin+2018, Zhdankin+2020, ComissoSironi2018, ComissoSironi2019, Comisso+2020, NattilaBeloborodov2020, Sobacchi+2021}. Particle acceleration proceeds in two stages \citep[e.g.][]{ComissoSironi2018, ComissoSironi2019}. First, particles experience an impulsive acceleration event that is powered by reconnection in large-scale current sheets. Since the reconnection electric field is nearly aligned with the local magnetic field, the distribution of the accelerated particles is strongly anisotropic (particles move nearly along the direction of the local magnetic field). Second, particles may be further accelerated by stochastic scattering off the turbulent magnetic fluctuations, similar to the original picture of \citet[][]{Fermi1949}. Stochastic acceleration is suppressed in fast cooling conditions since the acceleration timescale is comparable with the light crossing time of the system \citep[e.g.][]{NattilaBeloborodov2020, SobacchiLyubarsky2020, Zhdankin+2020, Sobacchi+2021}. Impulsive acceleration is practically unaffected by cooling since it operates on extremely short timescales.\footnote{Even though we focus on simulations of magnetically-dominated plasma turbulence, anisotropic particle distributions may be produced in any system where particle injection is governed by reconnection in the strong guide field regime, and where fast cooling prevents further particle energisation. This may happen in the non-linear stages of the kink instability \citep{Davelaar+2020} and of the Kelvin-Helmholtz instability \citep{Sironi+2021}.}

Motivated by these results, we study synchrotron-self-Compton emission from anisotropic particles.\footnote{In synchrotron-self-Compton emission, the synchrotron photons are IC scattered to higher energies by the non-thermal electrons within the jet. We neglect IC scattering off any photon field that is produced outside the jet.} The anisotropy has an important impact on the properties of the emitted radiation. Since particles move nearly along the direction of the local magnetic field, the rate of synchrotron cooling is strongly reduced. As a result, even in a magnetically dominated plasma, IC scattering can become the dominant cooling channel and shape the particle distribution function, in particular in the fast cooling regime. Then the radiation spectrum depends on the IC scattering regime. Particle cooling in the Thomson regime leads to soft synchrotron and IC spectra, $\nu F_\nu\propto\nu^{1/2}$, while cooling in the Klein-Nishina regime leads to hard synchrotron spectra, typically $\nu F_\nu\propto\nu$. Then the difference between blazars and GRBs could be explained if the IC scattering regime is different.

Several authors argued that hard GRB spectra may be due to IC cooling in the Klein-Nishina regime \citep[e.g.][]{Derishev+2001, Bosnjak+2009, Nakar+2009, Daigne+2011}. However, these authors did not consider the effect of particle anisotropy. Then IC cooling can have a strong effect on the particle distribution only in weakly magnetised plasmas. A basic point of the present paper is that strong particle anisotropy allows magnetically-dominated jets to emit in the IC dominated regime, with hard synchrotron spectra.

The paper is organised as follows. In Section \ref{sec:SSC} we discuss the general properties of our model. In Section \ref{sec:KN} we describe the emitted radiation spectrum. We refer the reader not interested in the technical details of the derivation to Tables \ref{tab:sync1}-\ref{tab:sync6}, where we summarise the properties of the radiation spectrum. In Section \ref{sec:astro} we discuss the astrophysical implications of our results.

\section{Physical model}
\label{sec:SSC}

We consider a turbulent plasma in the jet rest frame. The plasma may be  roughly described as a cloud of some density $n_{\rm e}$ and size $l\sim R/\Gamma$, where $\Gamma$ is the jet Lorentz factor at a radius $R$. The jet carries magnetic field $B$, and we assume that turbulence is strong, with fluctuations $\delta B\sim B$ on scale $l$. It is convenient to introduce the ``electron magnetisation'' parameter,
\begin{equation}
\label{eq:sigma}
\sigma_{\rm e} = \frac{U_{\rm B}}{n_{\rm e}m_{\rm e}c^2}\;,
\end{equation}
where $U_{\rm B}=B^2/8\pi$ is the magnetic energy density, $m_{\rm e}$ is the electron mass and $c$ is the speed of light.\footnote{If electrons are initially relativistically hot, the electron magnetisation in Eq. \eqref{eq:sigma} is usually normalised to the electron enthalpy density.} In electron-proton plasmas, the overall magnetisation (normalised with respect to the proton rest mass energy) is $\sigma =(m_{\rm e}/m_{\rm p})\sigma_{\rm e}$, where $m_{\rm p}$ is the proton mass. In pair plasmas, the overall magnetisation is $\sigma=\sigma_{\rm e}$. The magnetisation parameter $\sigma_{\rm e}$ is defined as the available magnetic energy per unit electron rest mass energy.

In the magnetically-dominated regime $\sigma\gg 1$, the magnetic energy is dissipated on a timescale
\begin{equation}
t_{\rm dyn}=\frac{l}{c}
\end{equation}
and generates a population of non-thermal particles \citep[e.g.][]{ComissoSironi2018, ComissoSironi2019}. The conservation of energy suggests that the impulsive acceleration by reconnection can be described as injection of energetic particles with Lorentz factors $\gamma\sim\sigma_{\rm e}$. We assume that the injected particles have pitch angles $\theta$ ($\theta$ is the angle between the particle velocity and the local magnetic field).

First principles simulations of magnetically-dominated turbulence mostly focused on pair plasmas. When the plasma has a proton component, we assume that impulsive acceleration by reconnection transfers a large fraction of the magnetic energy to the electrons. Our assumption is supported by studies of relativistic reconnection in electron-proton and electron-positron-proton plasmas \citep[e.g.][]{Ball+2018, Werner+2018, Petropoulou+2019}. Then the energised electrons have Lorentz factors $\gamma\sim\sigma_{\rm e}$, independent of the plasma composition.

The pitch angle remains constant while the particles cool since the synchrotron and IC photons are emitted nearly along the direction of the particle motion. We consider pitch angles $1/\gamma\lesssim\theta\lesssim 1$, so that the particle momentum transverse to the magnetic field is relativistic. The regime of extremely small pitch angles, $\theta\lesssim 1/\gamma$, has been discussed by \citet{LloydPetrosian2000, LloydPetrosian2002}.

\subsection{Electron energy distribution shaped by radiative cooling}

The particle injection rate per unit volume may be written as $(n_{\rm e}/t_{\rm dyn})\delta[\gamma-\sigma_{\rm e}]$, where $\delta[\ldots]$ is the Dirac delta function. Particles injected with $\gamma\sim\sigma_{\rm e}$ cool on a timescale $t_{\rm cool}\ll t_{\rm dyn}$ and form a steady distribution ${\rm d}n_{\rm e}/{\rm d}\gamma$ described by
\begin{equation}
\label{eq:cont}
\frac{{\rm d}}{{\rm d}\gamma}\left(\dot{\gamma}\frac{{\rm d} n_{\rm e}}{{\rm d}\gamma}\right) + \frac{n_{\rm e}}{t_{\rm dyn}}\delta\left[\gamma-\sigma_{\rm e}\right] = 0 \;,
\end{equation}
where $\dot{\gamma}$ is the rate of change of $\gamma$ due to radiative losses. We are neglecting the effect of pair creation via two-photon annihilation (we discuss this assumption in Section \ref{sec:pair}). Integrating Eq. \eqref{eq:cont}, one finds
\begin{equation}
\label{eq:n}
\frac{{\rm d} n_{\rm e}}{{\rm d}\gamma} = - \frac{n_{\rm e}}{t_{\rm dyn}\dot{\gamma}}\;.
\end{equation}
The particle distribution extends from $\gamma=\sigma_{\rm e}$ down to $\gamma=\gamma_{\rm cool}$, where $\gamma_{\rm cool}$ is defined by the condition that the particle cooling time is equal to the dynamical time, i.e. $t_{\rm dyn}=-\gamma/\dot{\gamma}$. By definition, in the fast cooling regime we have $\gamma_{\rm cool}\ll\sigma_{\rm e}$.

The particle loses energy via synchrotron and IC emission with rate $\dot{\gamma}m_{\rm e}c^2=-P_{\rm s}[\gamma]-P_{\rm IC}[\gamma]$. The synchrotron power is
\begin{equation}
\label{eq:Psync}
P_{\rm s}\left[\gamma\right] \simeq c\sigma_{\rm T}\theta^2U_{\rm B}\gamma^2\;,
\end{equation}
where $\sigma_{\rm T}$ is the Thomson cross section. We have taken into account that the synchrotron power is suppressed by a factor $\sin^2\theta\sim\theta^2$ when the energised particles have small pitch angles $\theta$. The IC power is
\begin{equation}
\label{eq:PIC}
P_{\rm IC}\left[\gamma\right] \simeq c\sigma_{\rm T}U_{\rm s, av}\gamma^2\;,
\end{equation}
where $U_{\rm s, av}$ is the ``available'' energy density of the synchrotron photons, i.e. the energy density of the synchrotron photons with energies smaller than the Klein-Nishina threshold,\footnote{When the spectrum of the target photons is described by a power law, i.e. $\nu F_\nu\propto \nu^\alpha$, IC losses are dominated by scattering of photons near the Klein-Nishina threshold for $\alpha\lesssim 3/2$ \citep[e.g.][]{Moderski+2005}. This condition is always satisfied by target synchrotron photons since $\alpha\lesssim 4/3$.} $m_{\rm e}c^2/\gamma$. The net cooling rate of the particle is then
\begin{equation}
\label{eq:dotg}
\dot{\gamma} = -\frac{\sigma_{\rm T}\gamma^2}{m_{\rm e}c}\left(\theta^2 U_{\rm B}+U_{\rm s, av}\right)\;.
\end{equation}
Note that we have assumed the synchrotron radiation field to be approximately isotropic. This assumption relies on the fact that the magnetic field is tangled on the scale of the emitting cloud, as expected for strong turbulence with $\delta B\sim B$.

\subsection{Synchrotron and IC radiation}

Electrons with Lorentz factor $\gamma$ radiate synchrotron photons of energy
\begin{equation}
\label{eq:esync}
\varepsilon_{\rm s}\left[\gamma\right] \simeq \theta\gamma^2\left(\frac{B}{B_{\rm q}}\right)m_{\rm e}c^2 \;,
\end{equation}
where $B_{\rm q}=m_{\rm e}^2c^3/\hbar e= 4.4 \times 10^{13}{\rm\; G}$ ($\hbar$ is the reduced Planck constant and $e$ is the electron charge), and $\theta$ is the pitch angle. Each particle radiates a synchrotron spectrum which peaks at $\varepsilon_{\rm s}$, has a slope of $4/3$ below the peak and an exponential cutoff above the peak. When the spectrum is convolved with an electron distribution, the net result is similar to what would be obtained if each particle emits all synchrotron photons with $\varepsilon_{\rm s}[\gamma]$. This approximation is used throughout this paper.

Most of the synchrotron energy is carried by photons with energy $\varepsilon_{\rm s,pk}=\varepsilon_{\rm s}[\sigma_{\rm e}]$. The photons with energies $\varepsilon_{\rm s,pk}$ are the main targets for IC scattering by an electron with Lorentz factor $\gamma$ as long as their scattering can occur in the Thomson regime, i.e. $\gamma\lesssim m_{\rm e} c^2/\varepsilon_{\rm s,pk}$. The resulting IC photons have energies $\varepsilon_{\rm IC}\simeq \gamma^2\epsilon_{\rm s,pk}$. In the opposite case, $\gamma\gtrsim m_{\rm e}c^2/\varepsilon_{\rm s,pk}$, the electron mainly scatters photons with $\epsilon_{\rm s}\simeq m_{\rm e}c^2/\gamma$ above which IC scattering is suppressed by the Klein-Nishina effects. Then, the IC photons carry a significant fraction of the electron energy $\gamma m_{\rm e}c^2$. The two regimes may be summarized as
\begin{equation}
\label{eq:eIC}
\varepsilon_{\rm IC}\left[\gamma\right] \simeq \min\left[\gamma^2\varepsilon_{\rm s,pk},\gamma m_{\rm e}c^2\right]\;.
\end{equation}
We define $U[\varepsilon]$ as the radiation energy density of photons per unit of $\log\epsilon$. Our goal is to evaluate $U_{\rm s}[\varepsilon_{\rm s}]$ and $U_{\rm IC}[\varepsilon_{\rm IC}]$ for the synchrotron and IC radiation. We assume that the current sheets are uniformly distributed throughout the plasma cloud. Then the radiation energy density is also approximately uniform. Since photons escape from the plasma cloud on a timescale $t_{\rm esc}=t_{\rm dyn}=l/c$, the energy density of radiation generated by electrons with Lorentz factors $\sim\gamma$ is $U_{\rm s}+U_{\rm IC}=\gamma ({\rm d} n_{\rm e}/{\rm d}\gamma)\;(P_{\rm s}+P_{\rm IC})\;t_{\rm dyn}$, which gives
\begin{equation}
U_{\rm s}+U_{\rm IC}=\left(\frac{\gamma}{\sigma_{\rm e}}\right)U_{\rm B}\;.
\end{equation}
Since we assumed that the magnetic energy converts to heat on the light crossing time $l/c$, and the heat quickly converts to radiation, energetic electrons with $\gamma\sim\sigma_{\rm e}$ emit a total radiation energy density $U_{\rm s}+U_{\rm IC}\sim U_{\rm B}$.

The synchrotron fraction $f_{\rm s}=U_{\rm s}/(U_{\rm s}+U_{\rm IC})=P_{\rm s}/(P_{\rm s}+P_{\rm IC})=\theta^2U_{\rm B}/(\theta^2 U_{\rm B}+U_{\rm s, av})$ gives
\begin{equation}
\label{eq:usync}
U_{\rm s}\left[\varepsilon_{\rm s}\right] = \frac{\theta^2 U_{\rm B}}{\theta^2 U_{\rm B}+U_{\rm s, av}}\left(\frac{\gamma}{\sigma_{\rm e}}\right) U_{\rm B} \;,
\end{equation}
where $\varepsilon_{\rm s}[\gamma]$ is given by Eq. \eqref{eq:esync}. When IC scattering occurs in the Thomson limit, and therefore $U_{\rm s, av}$ is independent of $\gamma$, from Eqs. \eqref{eq:esync} and \eqref{eq:usync} we recover the familiar result that $U_{\rm s}\propto\gamma\propto\varepsilon_{\rm s}^{1/2}$. The synchrotron spectrum reaches the peak at $\epsilon_{\rm s,pk}$ and is exponentially suppressed at $\epsilon_{\rm s}\gtrsim\epsilon_{\rm s,pk}$. The IC fraction $f_{\rm IC}=1-f_{\rm s}$ gives
\begin{equation}
\label{eq:uIC}
U_{\rm IC}\left[\varepsilon_{\rm IC}\right] = \frac{U_{\rm s, av}}{\theta^2 U_{\rm B}+U_{\rm s, av}}\left(\frac{\gamma}{\sigma_{\rm e}}\right) U_{\rm B} \;,
\end{equation}
where $\varepsilon_{\rm IC}[\gamma]$ is given by Eq. \eqref{eq:eIC}. When the IC scattering occurs in the Thomson limit, and therefore $U_{\rm s, av}$ is independent of $\gamma$, from Eqs. \eqref{eq:eIC} and \eqref{eq:uIC} we recover the familiar result that $U_{\rm IC}\propto\gamma\propto\varepsilon_{\rm IC}^{1/2}$.

\subsection{Electron cooling time and electron energy density}

Using Eqs. \eqref{eq:dotg} and \eqref{eq:usync}, the particle cooling time, $t_{\rm cool}[\gamma]=-\gamma/\dot{\gamma}$, can be conveniently expressed as
\begin{equation}
\label{eq:tcool}
t_{\rm cool}\left[\gamma\right] = \frac{U_{\rm s}}{U_{\rm B}} \frac{\sigma_{\rm e}}{\theta^2\gamma^2} \frac{t_{\rm dyn}}{\ell_{\rm B}} \;,
\end{equation}
where
\begin{equation}
\label{eq:lB}
\ell_{\rm B} = \frac{\sigma_{\rm T}U_{\rm B}t_{\rm dyn}}{m_{\rm e}c}
\end{equation}
is the magnetic compactness. When particles are isotropic, i.e. $\theta\sim 1$, cooling is dominated by synchrotron, and then $U_{\rm s}=(\gamma/\sigma_{\rm e})U_{\rm B}$. In this case, Eq. \eqref{eq:tcool} gives $t_{\rm cool}=t_{\rm dyn}/\gamma \ell_{\rm B}$. Within a dynamical time, electrons cool down to Lorentz factors $\gamma_{\rm cool}\simeq\max[1/\ell_{\rm B},1]$.

Using Eqs. \eqref{eq:n} and \eqref{eq:usync}, the energy density of the electrons with Lorentz factors $\sim\gamma$, $U_{\rm e}[\gamma]=(\gamma m_{\rm e} c^2)\times [\gamma ({\rm d}n_{\rm e}/{\rm d}\gamma)]$, can be expressed as
\begin{equation}
\label{eq:Ue}
U_{\rm e}\left[\gamma\right] = \frac{U_{\rm s}}{\gamma\theta^2 \ell_{\rm B}} = \frac{t_{\rm cool}}{t_{\rm dyn}}\left(\frac{\gamma}{\sigma_{\rm e}}\right)U_{\rm B} \;.
\end{equation}
The electron energy density is smaller than the energy density of the emitted radiation by the factor $t_{\rm cool}/t_{\rm dyn}\ll 1$. When a significant fraction of the magnetic energy is dissipated, the ratio of magnetic and electron energy densities is $U_{\rm B}/U_{\rm e}\sim t_{\rm dyn}/t_{\rm cool}^{\rm pk}$, where we have defined $t_{\rm cool}^{\rm pk}=t_{\rm cool}[\sigma_{\rm e}]$. 

\subsection{Optical depth for pair production}

IC photons of energy $\varepsilon_{\rm IC}$ may annihilate with target synchrotron photons of energy $\varepsilon_{\rm s}\gtrsim\varepsilon_{\rm thr}\simeq m_{\rm e}^2c^4/\varepsilon_{\rm IC}$, and create an electron-positron pair. There are targets for photon-photon collisions if $\varepsilon_{\rm thr}\lesssim\varepsilon_{\rm s,pk}$. This occurs if the IC photon was emitted in the Klein-Nishina regime with $\varepsilon_{\rm IC}\simeq \gamma m_{\rm e}c^2$. Since $\varepsilon_{\rm thr}\simeq m_{\rm e}c^2/\gamma$, the number density of the target synchrotron photons is $n_{\rm thr}\simeq (\gamma/m_{\rm e}c^2)U_{\rm s, av}$. The optical depth for pair production is $\tau_{\gamma\gamma}=\sigma_{\gamma\gamma}ct_{\rm dyn}n_{\rm thr}$ where $\sigma_{\gamma\gamma}$ depends on the spectrum of the target synchrotron photons, and is a fraction of $\sigma_{\rm T}$ \citep[e.g.][]{Svensson1987}.

The optical depth for pair production, $\tau_{\gamma\gamma}$, can be conveniently expressed as a function of $U_{\rm s}$ and $U_{\rm IC}$. When $U_{\rm s, av}\lesssim\theta^2U_{\rm B}$, from Eq. \eqref{eq:uIC} we may estimate $U_{\rm s, av}=\theta^2(\sigma_{\rm e}/\gamma)U_{\rm IC}$. Then the optical depth for pair production is
\begin{equation}
\tau_{\gamma\gamma}\left[\varepsilon_{\rm IC}\right] = \frac{\sigma_{\gamma\gamma}}{\sigma_{\rm T}} \frac{U_{\rm IC}}{U_{\rm B}} \theta^2 \sigma_{\rm e} \ell_{\rm B} \;.
\end{equation}
When instead $U_{\rm s, av}\gtrsim\theta^2U_{\rm B}$, from Eq. \eqref{eq:usync} we may estimate $U_{\rm s, av}=\theta^2(\gamma/\sigma_{\rm e})(U_{\rm B}/U_{\rm s})U_{\rm B}$. Then the optical depth is
\begin{equation}
\tau_{\gamma\gamma}\left[\varepsilon_{\rm IC}\right] = \frac{\sigma_{\gamma\gamma}}{\sigma_{\rm T}} \frac{U_{\rm B}}{U_{\rm s}}\frac{\theta^2\gamma^2}{\sigma_{\rm e}} \ell_{\rm B} \;.
\end{equation}
In this case, we see that $\tau_{\gamma\gamma}=(\sigma_{\gamma\gamma}/\sigma_{\rm T})(t_{\rm dyn}/t_{\rm cool})$, which may exceed unity in fast cooling conditions (however, note that $\sigma_{\gamma\gamma}\lesssim\sigma_{\rm T}$). When $\tau_{\gamma\gamma}\gtrsim 1$, a full Monte-Carlo simulation of the pair cascade may be needed to model the radiated spectrum (see \citet[][]{Beloborodov+2014}, where such simulations are performed for IC cascades in shock-heated plasma). In the present paper, we limit our analysis to the regime where synchrotron radiation from the secondary pairs does not dominate the emitted spectrum. This condition is further discussed in Section \ref{sec:pair}.

\section{Radiation spectrum}
\label{sec:KN}

In this section we describe the spectrum of synchrotron and IC radiation. We refer the reader not interested in the technical details of the derivation to Tables \ref{tab:sync1}-\ref{tab:sync6}, where we summarise our results. We use analytical estimates, neglecting numerical factors of order unity, to identify the possible emission regimes, and evaluate the spectral slope of the produced radiation in each regime.

The radiation spectrum depends on the electron distribution function ${\rm d}n_{\rm e}/{\rm d}\gamma$, which is shaped by cooling. Depending on the parameters of the problem (in particular the particle pitch angle $\theta$), the cooling may be dominated by synchrotron or IC losses, and the IC losses may occur in Thomson or Klein-Nishina regimes. Scattering occurs in the Thomson regime for electron Lorentz factors $\gamma\lesssim\gamma_{\rm KN}$, and in the Klein-Nishina regime for $\gamma\gtrsim\gamma_{\rm KN}$. The Lorentz factor $\gamma_{\rm KN}$ is determined by the condition $\gamma_{\rm KN}\varepsilon_{\rm s,pk}=m_{\rm e}c^2$, which gives
\begin{equation}
\gamma_{\rm KN} = \left(\frac{\theta_{\rm KN}}{\theta}\right)\sigma_{\rm e} \;,
\end{equation}
where we have defined
\begin{equation}
\label{eq:alphaKN}
\theta_{\rm KN} = \frac{1}{\sigma_{\rm e}^3}\left(\frac{B_{\rm q}}{B}\right) \;.
\end{equation}
When $\theta\lesssim\theta_{\rm KN}$ (and therefore $\sigma_{\rm e}\lesssim\gamma_{\rm KN}$), IC scattering occurs in the Thomson regime for all the particles in the system. When $\theta\gtrsim\theta_{\rm KN}$, IC scattering occurs in the Thomson regime for $\gamma\lesssim\gamma_{\rm KN}$, and in the Klein-Nishina regime for $\gamma_{\rm KN}\lesssim\gamma\lesssim\sigma_{\rm e}$.

\begin{figure}
\centering
\includegraphics[width=0.47\textwidth]{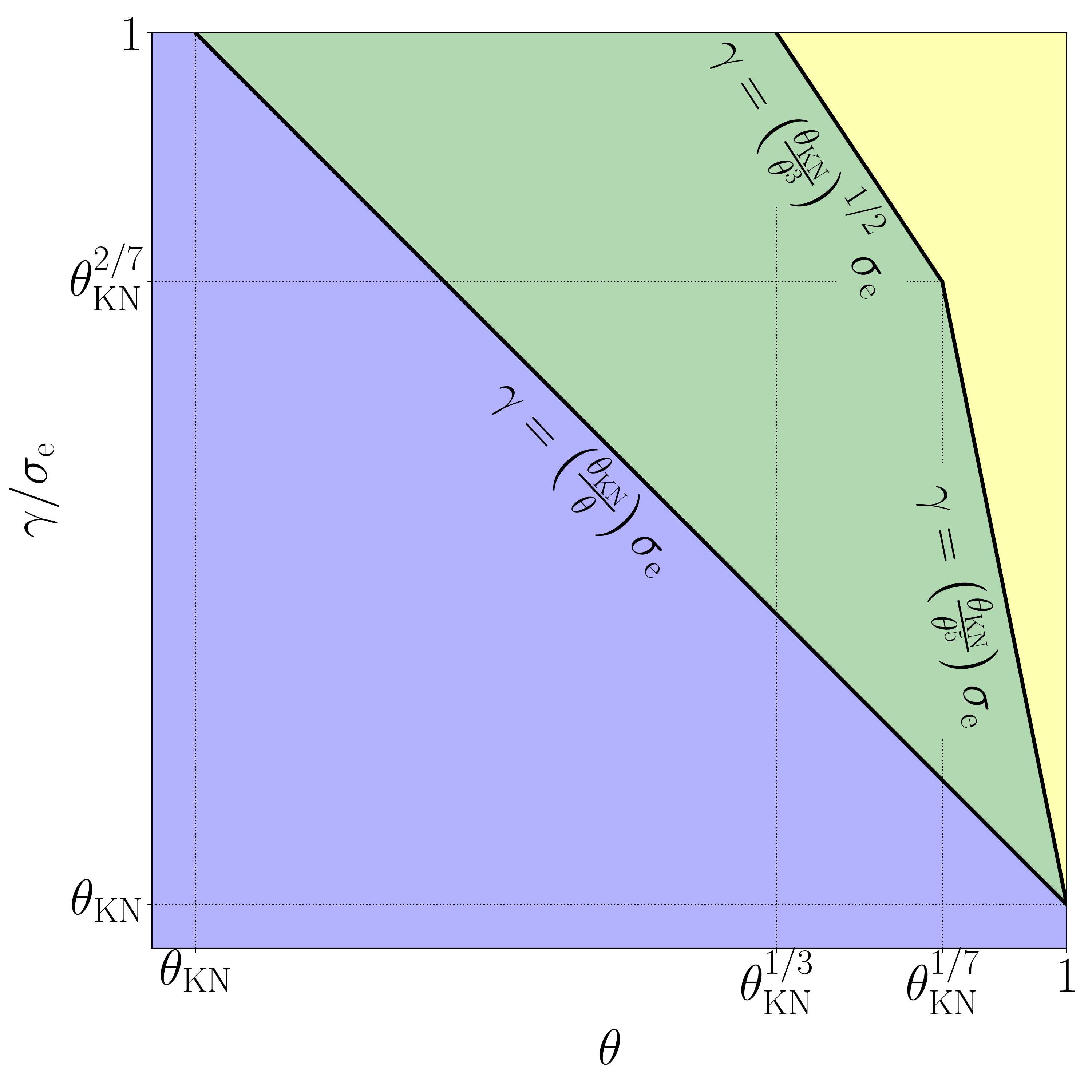}
\caption{Cooling regime for particles with Lorentz factor $\gamma$ and pitch angle $\theta$. Yellow: synchrotron dominated cooling. Green: IC dominated cooling (Klein-Nishina regime). Blue: IC dominated cooling (Thomson regime). Particles are injected with $\gamma\sim\sigma_{\rm e}$, and cool down at constant $\theta$.}
\label{fig:param}
\end{figure}

Electrons with $\gamma\lesssim\gamma_{\rm KN}$ IC scatter any synchrotron photons (with energies $\epsilon_{\rm s}$ up to the maximum $\epsilon_{\rm s,pk}$) in the Thomson regime. Since photons of energy $\varepsilon_{\rm s,pk}$ carry most of the synchrotron energy, $U_{\rm s, av}\simeq U_{\rm s}[\varepsilon_{\rm s,pk}]$, so Eqs. \eqref{eq:usync} and \eqref{eq:uIC} become
\begin{equation}
\label{eq:usyncT}
U_{\rm s}\left[\varepsilon_{\rm s}\right] = \frac{\theta^2 U_{\rm B}}{\theta^2 U_{\rm B}+U_{\rm s}\left[\varepsilon_{\rm s,pk}\right]}\left(\frac{\gamma}{\sigma_{\rm e}}\right) U_{\rm B}
\end{equation}
and
\begin{equation}
\label{eq:uICT}
U_{\rm IC}\left[\varepsilon_{\rm IC}\right] = \frac{U_{\rm s}\left[\varepsilon_{\rm s,pk}\right]}{\theta^2 U_{\rm B}+U_{\rm s}\left[\varepsilon_{\rm s,pk}\right]}\left(\frac{\gamma}{\sigma_{\rm e}}\right) U_{\rm B} \;.
\end{equation}
Electrons with $\gamma\gtrsim\gamma_{\rm KN}$ IC scatter photons of energy $\varepsilon_{\rm s,pk}$ in the Klein-Nishina regime. Since IC losses are dominated by scattering of photons near the Klein-Nishina threshold, the available synchrotron radiation is $U_{\rm s, av}=U_{\rm s}[\varepsilon_0]$, where we have defined
\begin{equation}
\varepsilon_0\left[\gamma\right] =\frac{m_{\rm e}c^2}{\gamma} \;.
\end{equation}
Eqs. \eqref{eq:usync} and \eqref{eq:uIC} now become
\begin{equation}
\label{eq:usyncKN}
U_{\rm s}\left[\varepsilon_{\rm s}\right] = \frac{\theta^2 U_{\rm B}}{\theta^2 U_{\rm B}+U_{\rm s}\left[\varepsilon_0\right]}\left(\frac{\gamma}{\sigma_{\rm e}}\right) U_{\rm B}
\end{equation}
and
\begin{equation}
\label{eq:uICKN}
U_{\rm IC}\left[\varepsilon_{\rm IC}\right] = \frac{U_{\rm s}\left[\varepsilon_0\right]}{\theta^2 U_{\rm B}+U_{\rm s}\left[\varepsilon_0\right]}\left(\frac{\gamma}{\sigma_{\rm e}}\right) U_{\rm B} \;.
\end{equation}
Particle cooling is dominated by synchrotron when $U_{\rm s}[\varepsilon_0]\lesssim\theta^2U_{\rm B}$ (in this case, we find that $U_{\rm IC}\lesssim U_{\rm s}\simeq (\gamma/\sigma_{\rm e})U_{\rm B}$). Particle cooling is dominated by IC when $U_{\rm s}[\varepsilon_0]\gtrsim\theta^2U_{\rm B}$ (in this case, we find that $U_{\rm s}\lesssim U_{\rm IC}\simeq (\gamma/\sigma_{\rm e})U_{\rm B}$). Since $U_{\rm s}[\varepsilon_0]$ is a decreasing function of $\gamma$, IC cooling generally dominates for small Lorentz factors, and synchrotron cooling dominates for large Lorentz factors. The cooling regimes for particles with Lorentz factor $\gamma$ and pitch angle $\theta$ are summarised in Figure \ref{fig:param}.

The Lorentz factor $\gamma_0$ of the electrons that emit synchrotron photons of energy $\varepsilon_0$ is determined by solving
\begin{equation}
\varepsilon_{\rm s}[\gamma_0]=\varepsilon_0\left[\gamma\right] \;,
\end{equation}
which gives
\begin{equation}
\label{eq:gamma0}
\gamma_0 \left[\gamma\right] = \left(\frac{\theta_{\rm KN}}{\theta}\right)^{1/2} \left(\frac{\sigma_{\rm e}}{\gamma}\right)^{1/2} \sigma_{\rm e} \;.
\end{equation}
In the formulas for radiation spectra given below it will be convenient to use the Lorentz factor $\gamma_{\ast}$ defined by $\gamma_{\ast}=\gamma_0[\gamma_{\ast}]$, which gives
\begin{equation}
\gamma_{\ast} = \left(\frac{\theta_{\rm KN}}{\theta}\right)^{1/3}\sigma_{\rm e} \;.
\end{equation}
Particles with $\gamma=\gamma_{\ast}$ IC scatter at the Klein-Nishina threshold the synchrotron photons that they themselves emit. For these particles $U_{\rm s}[\varepsilon_0]=U_{\rm s}[\varepsilon_{\ast}]$, where we have defined $\varepsilon_{\ast}=\varepsilon_{\rm s}[\gamma_{\ast}]$. Then substituting $\gamma=\gamma_{\ast}$ into Eq. \eqref{eq:usyncKN} gives
\begin{equation}
\label{eq:uast}
U_{\rm s}\left[\varepsilon_{\ast}\right]= \frac{2\left(\theta_{\rm KN}/\theta\right)^{1/3}}{1+\sqrt{1+ 4\left(\theta_{\rm KN}/\theta^7\right)^{1/3}}} U_{\rm B} \;.
\end{equation}
Eqs. \eqref{eq:usyncT}-\eqref{eq:uICT} and \eqref{eq:usyncKN}-\eqref{eq:uICKN} can be used to find the radiation spectrum in all possible regimes.

\begin{table*}
\begin{center}
\begin{tabular}{| c |}
\hline
$\gamma\lesssim \sigma_{\rm e}$ \\ [0.5ex] 
\hline\hline
$U_{\rm s}\left[\varepsilon_{\rm s}\right] = \theta \left(\frac{\gamma}{\sigma_{\rm e}}\right) U_{\rm B}$ \\ [0.5ex]
\hline
$U_{\rm IC}\left[\varepsilon_{\rm IC}\right] = \left(\frac{\gamma}{\sigma_{\rm e}}\right) U_{\rm B}$ \\ [0.5ex] 
\hline
$t_{\rm cool}\left[\gamma\right] = \left(\frac{1}{\theta}\right) \left(\frac{1}{\gamma}\right) \frac{t_{\rm dyn}}{\ell_{\rm B}}$ \\ [0.5ex]
\hline
$\tau_{\gamma\gamma}\left[\varepsilon_{\rm IC}\right]=0$ \\ [0.5ex]
\hline
\end{tabular}
\caption{\label{tab:sync1} Energy density of the synchrotron and IC photons emitted by particles with Lorentz factor $\gamma$, particle cooling time, and optical depth for pair production, for pitch angles $\theta \lesssim\theta_{\rm KN}$ (see Section \ref{sec:pitchT}). The synchrotron photon energy is $\varepsilon_{\rm s}\simeq (\theta/\theta_{\rm KN})(\gamma^2/\sigma_{\rm e}^3)m_{\rm e}c^2$. The IC photon energy is $\varepsilon_{\rm IC}\simeq (\theta/\theta_{\rm KN})(\gamma^2/\sigma_{\rm e})m_{\rm e}c^2$. We assume fast cooling conditions, i.e. $t_{\rm cool}\lesssim t_{\rm dyn}$. The magnetic compactness $\ell_{\rm B}$ and the critical pitch angle $\theta_{\rm KN}$ are defined in Eqs. \eqref{eq:lB} and \eqref{eq:alphaKN} respectively.}
\end{center}
\end{table*}

\begin{table*}
\begin{center}
\begin{tabular}{| c | c | }
\hline
$\gamma\lesssim \left(\frac{\theta_{\rm KN}}{\theta}\right)\sigma_{\rm e}$ & $\left(\frac{\theta_{\rm KN}}{\theta}\right)\sigma_{\rm e} \lesssim \gamma\lesssim \sigma_{\rm e}$ \\ [0.5ex] 
\hline\hline
$U_{\rm s}\left[\varepsilon_{\rm s}\right] = \left(\theta\theta_{\rm KN}\right)^{1/2} \left(\frac{\gamma}{\sigma_{\rm e}}\right) U_{\rm B}$ & $U_{\rm s}\left[\varepsilon_{\rm s}\right] = \left(\frac{\theta^3}{\theta_{\rm KN}}\right)^{1/2}\left(\frac{\gamma}{\sigma_{\rm e}}\right)^2 U_{\rm B}$ \\ [0.5ex]
\hline
\multicolumn{2}{|c|}{$U_{\rm IC}\left[\varepsilon_{\rm IC}\right] = \left(\frac{\gamma}{\sigma_{\rm e}}\right) U_{\rm B}$} \\ [0.5ex] 
\hline
$t_{\rm cool}\left[\gamma\right]= \left(\frac{\theta_{\rm KN}}{\theta^3}\right)^{1/2} \left(\frac{1}{\gamma}\right) \frac{t_{\rm dyn}}{\ell_{\rm B}}$ & $t_{\rm cool}\left[\gamma\right]= \left(\frac{1}{\theta\theta_{\rm KN}}\right)^{1/2} \left(\frac{1}{\sigma_{\rm e}}\right) \frac{t_{\rm dyn}}{\ell_{\rm B}}$ \\ [0.5ex]
\hline
$\tau_{\gamma\gamma}\left[\varepsilon_{\rm IC}\right]=0$ & $\tau_{\gamma\gamma}\left[\varepsilon_{\rm IC}\right]=\left(\frac{\sigma_{\gamma\gamma}}{\sigma_{\rm T}}\right) \left(\theta\theta_{\rm KN}\right)^{1/2} \sigma_{\rm e} \ell_{\rm B}$ \\ [0.5ex]
\hline
\end{tabular}
\caption{\label{tab:sync2} Same as Table \ref{tab:sync1}, for pitch angles $\theta_{\rm KN}\lesssim\theta\lesssim \theta_{\rm KN}^{1/3}$ (see Section \ref{sec:pitchKNsmall}). The synchrotron photon energy is $\varepsilon_{\rm s}\simeq (\theta/\theta_{\rm KN})(\gamma^2/\sigma_{\rm e}^3)m_{\rm e}c^2$. The IC photon energy is $\varepsilon_{\rm IC}\simeq (\theta/\theta_{\rm KN})(\gamma^2/\sigma_{\rm e})m_{\rm e}c^2$ if $\gamma\lesssim(\theta_{\rm KN}/\theta)\sigma_{\rm e}$, and $\varepsilon_{\rm IC}\simeq \gamma m_{\rm e}c^2$ if $\gamma\gtrsim(\theta_{\rm KN}/\theta)\sigma_{\rm e}$.}
\end{center}
\end{table*}

\begin{table*}
\begin{center}
\begin{tabular}{| c | c | c | c|}
\hline
$\gamma\lesssim \left(\frac{\theta_{\rm KN}}{\theta}\right)\sigma_{\rm e}$ & $\left(\frac{\theta_{\rm KN}}{\theta}\right)\sigma_{\rm e}\lesssim\gamma\lesssim\theta^2\sigma_{\rm e}$ & $\theta^2 \sigma_{\rm e}\lesssim \gamma\lesssim \left(\frac{\theta_{\rm KN}}{\theta^3}\right)^{1/2}\sigma_{\rm e}$ & $\left(\frac{\theta_{\rm KN}}{\theta^3}\right)^{1/2}\sigma_{\rm e}\lesssim\gamma\lesssim\sigma_{\rm e}$ \\ [0.5ex] 
\hline\hline
$U_{\rm s}\left[\varepsilon_{\rm s}\right] = \theta^2 \left(\frac{\gamma}{\sigma_{\rm e}}\right) U_{\rm B}$ & $U_{\rm s}\left[\varepsilon_{\rm s}\right] = \left(\frac{\theta^5}{\theta_{\rm KN}}\right)^{1/2}\left(\frac{\gamma}{\sigma_{\rm e}}\right)^{3/2}U_{\rm B}$ & $U_{\rm s}\left[\varepsilon_{\rm s}\right] = \left(\frac{\theta^3}{\theta_{\rm KN}}\right)^{1/2} \left(\frac{\gamma}{\sigma_{\rm e}}\right)^2 U_{\rm B}$ & $U_{\rm s}\left[\varepsilon_{\rm s}\right] = \left(\frac{\gamma}{\sigma_{\rm e}}\right)U_{\rm B}$ \\ [0.5ex]
\hline
\multicolumn{3}{|c|}{$U_{\rm IC}\left[\varepsilon_{\rm IC}\right] = \left(\frac{\gamma}{\sigma_{\rm e}}\right) U_{\rm B}$} & $U_{\rm IC}\left[\varepsilon_{\rm IC}\right] = \left(\frac{\theta_{\rm KN}}{\theta^3}\right)^{1/2} U_{\rm B}$ \\ [0.5ex]
\hline
$t_{\rm cool}\left[\gamma\right]= \left(\frac{1}{\gamma}\right) \frac{t_{\rm dyn}}{\ell_{\rm B}}$ & $t_{\rm cool}\left[\gamma\right]=\left(\frac{\theta}{\theta_{\rm KN}}\right)^{1/2}\left(\frac{1}{\gamma\sigma_{\rm e}}\right)^{1/2} \frac{t_{\rm dyn}}{\ell_{\rm B}}$ & $t_{\rm cool}\left[\gamma\right]=\left(\frac{1}{\theta\theta_{\rm KN}}\right)^{1/2} \left(\frac{1}{\sigma_{\rm e}}\right) \frac{t_{\rm dyn}}{\ell_{\rm B}}$ & $t_{\rm cool}\left[\gamma\right]=\left(\frac{1}{\theta}\right)^2 \left(\frac{1}{\gamma}\right) \frac{t_{\rm dyn}}{\ell_{\rm B}}$ \\ [0.5ex] 
\hline
$\tau_{\gamma\gamma}\left[\varepsilon_{\rm IC}\right]=0$ & $\tau_{\gamma\gamma}\left[\varepsilon_{\rm IC}\right]=\left(\frac{\sigma_{\gamma\gamma}}{\sigma_{\rm T}}\right) \left(\frac{\theta_{\rm KN}}{\theta}\right)^{1/2} \left(\gamma\sigma_{\rm e}\right)^{1/2} \ell_{\rm B}$ & \multicolumn{2}{c|}{$\tau_{\gamma\gamma}\left[\varepsilon_{\rm IC}\right]=\left(\frac{\sigma_{\gamma\gamma}}{\sigma_{\rm T}}\right) \left(\theta\theta_{\rm KN}\right)^{1/2} \sigma_{\rm e} \ell_{\rm B}$} \\ [0.5ex]
\hline
\end{tabular}
\caption{\label{tab:sync3} Same as Table \ref{tab:sync1}, for pitch angles $\theta_{\rm KN}^{1/3}\lesssim\theta\lesssim \theta_{\rm KN}^{1/5}$ (see Section \ref{sec:pitchKNint}). The synchrotron photon energy is $\varepsilon_{\rm s}\simeq (\theta/\theta_{\rm KN})(\gamma^2/\sigma_{\rm e}^3)m_{\rm e}c^2$. The IC photon energy is $\varepsilon_{\rm IC}\simeq (\theta/\theta_{\rm KN})(\gamma^2/\sigma_{\rm e})m_{\rm e}c^2$ if $\gamma\lesssim(\theta_{\rm KN}/\theta)\sigma_{\rm e}$, and $\varepsilon_{\rm IC}\simeq \gamma m_{\rm e}c^2$ if $\gamma\gtrsim(\theta_{\rm KN}/\theta)\sigma_{\rm e}$.}
\end{center}
\end{table*}

\begin{table*}
\begin{center}
\begin{tabular}{| c | c | c |}
\hline
$\gamma\lesssim \left(\frac{\theta_{\rm KN}}{\theta}\right)\sigma_{\rm e}$ & $\left(\frac{\theta_{\rm KN}}{\theta}\right)\sigma_{\rm e}\lesssim\gamma\lesssim\theta^2\sigma_{\rm e}$ & $\theta^2 \sigma_{\rm e}\lesssim \gamma\lesssim \left(\frac{\theta_{\rm KN}}{\theta^3}\right)^{1/2}\sigma_{\rm e}$ \\ [0.5ex] 
\hline\hline
$U_{\rm s}\left[\varepsilon_{\rm s}\right] = \theta^2 \left(\frac{\gamma}{\sigma_{\rm e}}\right) U_{\rm B}$ & $U_{\rm s}\left[\varepsilon_{\rm s}\right] = \left(\frac{\theta^5}{\theta_{\rm KN}}\right)^{1/2}\left(\frac{\gamma}{\sigma_{\rm e}}\right)^{3/2}U_{\rm B}$ & $U_{\rm s}\left[\varepsilon_{\rm s}\right] = \left(\frac{\theta^3}{\theta_{\rm KN}}\right)^{1/2} \left(\frac{\gamma}{\sigma_{\rm e}}\right)^2 U_{\rm B}$ \\ [0.5ex]
\hline
\multicolumn{3}{|c|}{$U_{\rm IC}\left[\varepsilon_{\rm IC}\right] = \left(\frac{\gamma}{\sigma_{\rm e}}\right) U_{\rm B}$} \\ [0.5ex]
\hline
$t_{\rm cool}\left[\gamma\right]= \left(\frac{1}{\gamma}\right) \frac{t_{\rm dyn}}{\ell_{\rm B}}$ & $t_{\rm cool}\left[\gamma\right]=\left(\frac{\theta}{\theta_{\rm KN}}\right)^{1/2}\left(\frac{1}{\gamma\sigma_{\rm e}}\right)^{1/2} \frac{t_{\rm dyn}}{\ell_{\rm B}}$ & $t_{\rm cool}\left[\gamma\right]=\left(\frac{1}{\theta\theta_{\rm KN}}\right)^{1/2} \left(\frac{1}{\sigma_{\rm e}}\right)\frac{t_{\rm dyn}}{\ell_{\rm B}}$ \\ [0.5ex] 
\hline
$\tau_{\gamma\gamma}\left[\varepsilon_{\rm IC}\right]=0$ & $\tau_{\gamma\gamma}\left[\varepsilon_{\rm IC}\right]=\left(\frac{\sigma_{\gamma\gamma}}{\sigma_{\rm T}}\right) \left(\frac{\theta_{\rm KN}}{\theta}\right)^{1/2} \left(\gamma\sigma_{\rm e}\right)^{1/2} \ell_{\rm B}$ & $\tau_{\gamma\gamma}\left[\varepsilon_{\rm IC}\right]=\left(\frac{\sigma_{\gamma\gamma}}{\sigma_{\rm T}}\right) \left(\theta\theta_{\rm KN}\right)^{1/2} \sigma_{\rm e} \ell_{\rm B}$ \\ [0.5ex]
\hline
\end{tabular}
\\ [1.0ex]
\begin{tabular}{| c | c |}
\hline
$\left(\frac{\theta_{\rm KN}}{\theta^3}\right)^{1/2}\sigma_{\rm e}\lesssim\gamma\lesssim \left(\frac{\theta_{\rm KN}}{\theta^5}\right)\sigma_{\rm e}$ & $\left(\frac{\theta_{\rm KN}}{\theta^5}\right)\sigma_{\rm e}\lesssim\gamma\lesssim\sigma_{\rm e}$ \\ [0.5ex] 
\hline\hline
\multicolumn{2}{|c|}{$U_{\rm s}\left[\varepsilon_{\rm s}\right] = \left(\frac{\gamma}{\sigma_{\rm e}}\right)U_{\rm B}$} \\ [0.5ex]
\hline
$U_{\rm IC}\left[\varepsilon_{\rm IC}\right] = \left(\frac{\theta_{\rm KN}}{\theta^3}\right)^{1/2} U_{\rm B}$ & $U_{\rm IC}\left[\varepsilon_{\rm IC}\right] = \left(\frac{\theta_{\rm KN}}{\theta}\right)^{1/4} \left(\frac{\gamma}{\sigma_{\rm e}}\right)^{1/4} U_{\rm B}$ \\ [0.5ex]
\hline
\multicolumn{2}{|c|}{$t_{\rm cool}\left[\gamma\right]=\left(\frac{1}{\theta}\right)^2 \left(\frac{1}{\gamma}\right) \frac{t_{\rm dyn}}{\ell_{\rm B}}$} \\ [0.5ex] 
\hline
$\tau_{\gamma\gamma}\left[\varepsilon_{\rm IC}\right]=\left(\frac{\sigma_{\gamma\gamma}}{\sigma_{\rm T}}\right) \left(\theta\theta_{\rm KN}\right)^{1/2} \sigma_{\rm e} \ell_{\rm B}$ & $\tau_{\gamma\gamma}\left[\varepsilon_{\rm IC}\right]=\left(\frac{\sigma_{\gamma\gamma}}{\sigma_{\rm T}}\right) \left(\theta^7\theta_{\rm KN}\right)^{1/4} \left(\gamma\sigma_{\rm e}^3\right)^{1/4} \ell_{\rm B}$ \\ [0.5ex]
\hline
\end{tabular}
\caption{\label{tab:sync4} Same as Table \ref{tab:sync1}, for pitch angles $\theta_{\rm KN}^{1/5}\lesssim\theta\lesssim \theta_{\rm KN}^{1/7}$ (see Section \ref{sec:pitchKNint}). The synchrotron photon energy is $\varepsilon_{\rm s}\simeq (\theta/\theta_{\rm KN})(\gamma^2/\sigma_{\rm e}^3)m_{\rm e}c^2$. The IC photon energy is $\varepsilon_{\rm IC}\simeq (\theta/\theta_{\rm KN})(\gamma^2/\sigma_{\rm e})m_{\rm e}c^2$ if $\gamma\lesssim(\theta_{\rm KN}/\theta)\sigma_{\rm e}$, and $\varepsilon_{\rm IC}\simeq \gamma m_{\rm e}c^2$ if $\gamma\gtrsim(\theta_{\rm KN}/\theta)\sigma_{\rm e}$.}
\end{center}
\end{table*}

\begin{table*}
\begin{center}
\begin{tabular}{| c | c | c | c | }
\hline
$\gamma\lesssim \left(\frac{\theta_{\rm KN}}{\theta}\right)\sigma_{\rm e}$ & $\left(\frac{\theta_{\rm KN}}{\theta}\right)\sigma_{\rm e}\lesssim \gamma\lesssim \left(\frac{\theta_{\rm KN}}{\theta^5}\right)\sigma_{\rm e}$ & $\left(\frac{\theta_{\rm KN}}{\theta^5}\right)\sigma_{\rm e}\lesssim\gamma\lesssim\left(\frac{\theta^9}{\theta_{\rm KN}}\right)\sigma_{\rm e}$ & $\left(\frac{\theta^9}{\theta_{\rm KN}}\right)\sigma_{\rm e}\lesssim\gamma\lesssim\sigma_{\rm e}$ \\ [0.5ex] 
\hline\hline
$U_{\rm s}\left[\varepsilon_{\rm s}\right] = \theta^2 \left(\frac{\gamma}{\sigma_{\rm e}}\right) U_{\rm B}$ & $U_{\rm s}\left[\varepsilon_{\rm s}\right] = \left(\frac{\theta^5}{\theta_{\rm KN}}\right)^{1/2}\left(\frac{\gamma}{\sigma_{\rm e}}\right)^{3/2}U_{\rm B}$ & \multicolumn{2}{c|}{$U_{\rm s}\left[\varepsilon_{\rm s}\right] = \left(\frac{\gamma}{\sigma_{\rm e}}\right)U_{\rm B}$} \\ [0.5ex] 
\hline
\multicolumn{2}{|c|}{$U_{\rm IC}\left[\varepsilon_{\rm IC}\right] = \left(\frac{\gamma}{\sigma_{\rm e}}\right) U_{\rm B}$} & $U_{\rm IC}\left[\varepsilon_{\rm IC}\right] = \left(\frac{\theta_{\rm KN}}{\theta^5}\right)^{1/2} \left(\frac{\gamma}{\sigma_{\rm e}}\right)^{1/2} U_{\rm B}$ & $U_{\rm IC}\left[\varepsilon_{\rm IC}\right] = \left(\frac{\theta_{\rm KN}}{\theta}\right)^{1/4} \left(\frac{\gamma}{\sigma_{\rm e}}\right)^{1/4} U_{\rm B}$ \\ [0.5ex] 
\hline
$t_{\rm cool}\left[\gamma\right]=\left(\frac{1}{\gamma}\right) \frac{t_{\rm dyn}}{\ell_{\rm B}}$ & $t_{\rm cool}\left[\gamma\right]=\left(\frac{\theta}{\theta_{\rm KN}}\right)^{1/2}\left(\frac{1}{\gamma\sigma_{\rm e}}\right)^{1/2} \frac{t_{\rm dyn}}{\ell_{\rm B}}$ & \multicolumn{2}{c|}{$t_{\rm cool}\left[\gamma\right]=\left(\frac{1}{\theta}\right)^2\left(\frac{1}{\gamma}\right) \frac{t_{\rm dyn}}{\ell_{\rm B}}$} \\ [0.5ex] 
\hline
$\tau_{\gamma\gamma}\left[\varepsilon_{\rm IC}\right]=0$ & \multicolumn{2}{c|}{$\tau_{\gamma\gamma}\left[\varepsilon_{\rm IC}\right]=\left(\frac{\sigma_{\gamma\gamma}}{\sigma_{\rm T}}\right) \left(\frac{\theta_{\rm KN}}{\theta}\right)^{1/2} \left(\gamma\sigma_{\rm e}\right)^{1/2} \ell_{\rm B}$} & $\tau_{\gamma\gamma}\left[\varepsilon_{\rm IC}\right]=\left(\frac{\sigma_{\gamma\gamma}}{\sigma_{\rm T}}\right) \left(\theta^7\theta_{\rm KN}\right)^{1/4} \left(\gamma\sigma_{\rm e}^3\right)^{1/4} \ell_{\rm B}$ \\ [0.5ex]
\hline
\end{tabular}
\caption{\label{tab:sync5} Same as Table \ref{tab:sync1}, for pitch angles $\theta_{\rm KN}^{1/7}\lesssim\theta\lesssim \theta_{\rm KN}^{1/9}$ (see Section \ref{sec:pitchKNlarge}). The synchrotron photon energy is $\varepsilon_{\rm s}\simeq (\theta/\theta_{\rm KN})(\gamma^2/\sigma_{\rm e}^3)m_{\rm e}c^2$. The IC photon energy is $\varepsilon_{\rm IC}\simeq (\theta/\theta_{\rm KN})(\gamma^2/\sigma_{\rm e})m_{\rm e}c^2$ if $\gamma\lesssim(\theta_{\rm KN}/\theta)\sigma_{\rm e}$, and $\varepsilon_{\rm IC}\simeq \gamma m_{\rm e}c^2$ if $\gamma\gtrsim(\theta_{\rm KN}/\theta)\sigma_{\rm e}$.}
\end{center}
\end{table*}

\begin{table*}
\begin{center}
\begin{tabular}{| c | c | c | }
\hline
$\gamma\lesssim \left(\frac{\theta_{\rm KN}}{\theta}\right)\sigma_{\rm e}$ & $\left(\frac{\theta_{\rm KN}}{\theta}\right)\sigma_{\rm e}\lesssim \gamma\lesssim \left(\frac{\theta_{\rm KN}}{\theta^5}\right)\sigma_{\rm e}$ & $\left(\frac{\theta_{\rm KN}}{\theta^5}\right)\sigma_{\rm e}\lesssim\gamma\lesssim\sigma_{\rm e}$ \\ [0.5ex] 
\hline\hline
$U_{\rm s}\left[\varepsilon_{\rm s}\right] = \theta^2 \left(\frac{\gamma}{\sigma_{\rm e}}\right) U_{\rm B}$ & $U_{\rm s}\left[\varepsilon_{\rm s}\right] = \left(\frac{\theta^5}{\theta_{\rm KN}}\right)^{1/2}\left(\frac{\gamma}{\sigma_{\rm e}}\right)^{3/2}U_{\rm B}$ & $U_{\rm s}\left[\varepsilon_{\rm s}\right] = \left(\frac{\gamma}{\sigma_{\rm e}}\right)U_{\rm B}$ \\ [0.5ex] 
\hline
\multicolumn{2}{|c|}{$U_{\rm IC}\left[\varepsilon_{\rm IC}\right] = \left(\frac{\gamma}{\sigma_{\rm e}}\right) U_{\rm B}$} & $U_{\rm IC}\left[\varepsilon_{\rm IC}\right] = \left(\frac{\theta_{\rm KN}}{\theta^5}\right)^{1/2} \left(\frac{\gamma}{\sigma_{\rm e}}\right)^{1/2} U_{\rm B}$ \\ [0.5ex] 
\hline
$t_{\rm cool}\left[\gamma\right]=\left(\frac{1}{\gamma}\right)\frac{t_{\rm dyn}}{\ell_{\rm B}}$ & $t_{\rm cool}\left[\gamma\right]=\left(\frac{\theta}{\theta_{\rm KN}}\right)^{1/2}\left(\frac{1}{\gamma\sigma_{\rm e}}\right)^{1/2}\frac{t_{\rm dyn}}{\ell_{\rm B}}$ & $t_{\rm cool}\left[\gamma\right]=\left(\frac{1}{\theta}\right)^2\left(\frac{1}{\gamma}\right)\frac{t_{\rm dyn}}{\ell_{\rm B}}$ \\ [0.5ex] 
\hline
$\tau_{\gamma\gamma}\left[\varepsilon_{\rm IC}\right]=0$ & \multicolumn{2}{c|}{$\tau_{\gamma\gamma}\left[\varepsilon_{\rm IC}\right]=\left(\frac{\sigma_{\gamma\gamma}}{\sigma_{\rm T}}\right) \left(\frac{\theta_{\rm KN}}{\theta}\right)^{1/2} \left(\gamma\sigma_{\rm e}\right)^{1/2} \ell_{\rm B}$} \\ [0.5ex]
\hline
\end{tabular}
\caption{\label{tab:sync6} Same as Table \ref{tab:sync1}, for pitch angles $\theta_{\rm KN}^{1/9}\lesssim\theta\lesssim 1$ (see Section \ref{sec:pitchKNlarge}). The synchrotron photon energy is $\varepsilon_{\rm s}\simeq (\theta/\theta_{\rm KN})(\gamma^2/\sigma_{\rm e}^3)m_{\rm e}c^2$. The IC photon energy is $\varepsilon_{\rm IC}\simeq (\theta/\theta_{\rm KN})(\gamma^2/\sigma_{\rm e})m_{\rm e}c^2$ if $\gamma\lesssim(\theta_{\rm KN}/\theta)\sigma_{\rm e}$, and $\varepsilon_{\rm IC}\simeq \gamma m_{\rm e}c^2$ if $\gamma\gtrsim(\theta_{\rm KN}/\theta)\sigma_{\rm e}$.}
\end{center}
\end{table*}

\subsection{Thomson regime}
\label{sec:pitchT}

The synchrotron and IC spectra are easily determined in the Thomson regime,
\begin{equation}
\theta\lesssim \theta_{\rm KN}\;.
\end{equation}
In this regime, even the most energetic particles in the system, with Lorentz factors $\gamma=\sigma_{\rm e}$, IC scatter photons of energy $\varepsilon_{\rm s,pk}$ in the Thomson regime.

Electrons with $\gamma=\sigma_{\rm e}$ emit synchrotron photons of energy $\varepsilon_{\rm s,pk}$, and IC photons of energy $\varepsilon_{\rm IC,pk}=\sigma_{\rm e}^2\varepsilon_{\rm s,pk}$. Substituting $\gamma=\sigma_{\rm e}$ into Eq. \eqref{eq:usyncT} we find that $U_{\rm s}[\varepsilon_{\rm s,pk}]=[(\sqrt{\theta^4+4\theta^2}-\theta^2)/2]U_{\rm B}$. Then $U_{\rm IC}[\varepsilon_{\rm IC,pk}]=U_{\rm B}-U_{\rm s}[\varepsilon_{\rm s,pk}]=[(2+\theta^2-\sqrt{\theta^4+4\theta^2})/2]U_{\rm B}$.

When particles are isotropic, i.e. $\theta\sim 1$, the above expressions give $U_{\rm s}[\varepsilon_{\rm s,pk}]\sim U_{\rm IC}[\varepsilon_{\rm IC,pk}]\sim U_{\rm B}$. When particles are strongly anisotropic, i.e. $\theta\ll 1$, the expressions give $U_{\rm s}[\varepsilon_{\rm s,pk}]\sim \theta U_{\rm B}$, and $U_{\rm IC}[\varepsilon_{\rm IC,pk}]\sim U_{\rm B}$. A simple approximation is then $U_{\rm s}[\varepsilon_{\rm s,pk}]=\theta U_{\rm B}$ and $U_{\rm IC}[\varepsilon_{\rm IC,pk}]=U_{\rm B}$. Then the synchrotron spectrum is
\begin{equation}
\label{eq:specsyncT}
U_{\rm s}\left[\varepsilon_{\rm s}\right]=\theta\left(\frac{\gamma}{\sigma_{\rm e}}\right)U_{\rm B}\propto\varepsilon_{\rm s}^{1/2}\;,
\end{equation}
and the IC spectrum is
\begin{equation}
\label{eq:specICT}
U_{\rm IC}\left[\varepsilon_{\rm IC}\right]=\left(\frac{\gamma}{\sigma_{\rm e}}\right)U_{\rm B}\propto\varepsilon_{\rm IC}^{1/2}
\end{equation}
for all Lorentz factors $\gamma\lesssim\sigma_{\rm e}$. Note that $U_{\rm s}/U_{\rm IC}\sim\theta$. Our results are summarised in Table \ref{tab:sync1}.

\subsection{Klein-Nishina regime}

\subsubsection{Large pitch angles}
\label{sec:pitchKNlarge}

First we consider the regime of large pitch angles,
\begin{equation}
\theta_{\rm KN}^{1/7}\lesssim\theta\lesssim 1\;.
\end{equation}
In this regime synchrotron dominates the cooling of the most energetic particles. Eq. \eqref{eq:uast} gives $U_{\rm s}[\varepsilon_{\ast}]=(\theta_{\rm KN}/\theta)^{1/3}U_{\rm B}$. Then particles with $\gamma=\gamma_{\ast}$ have $U_{\rm s}[\varepsilon_0]=(\theta_{\rm KN}/\theta)^{1/3}U_{\rm B}\lesssim\theta^2U_{\rm B}$. Since $U_{\rm s}[\varepsilon_0]$ is a decreasing function of $\gamma$, also particles with $\gamma=\sigma_{\rm e}$ have $U_{\rm s}[\varepsilon_0]\lesssim\theta^2U_{\rm B}$. Then $U_{\rm s}[\varepsilon_{\rm s,pk}]=U_{\rm B}$, where $\varepsilon_{\rm s,pk}=\varepsilon_{\rm s}[\sigma_{\rm e}]$. 

The synchrotron spectrum has two breaks. A low energy break occurs when IC cooling transitions from the Thomson regime (for $\gamma\lesssim\gamma_{\rm KN}$) to the Klein-Nishina regime (for $\gamma\gtrsim\gamma_{\rm KN}$). A high energy break occurs when cooling transitions from the IC dominated regime (for $\gamma\lesssim\gamma_{\rm b}$) to the synchrotron dominated regime (for $\gamma\gtrsim\gamma_{\rm b}$). The Lorentz factor $\gamma_{\rm b}$ is determined by the condition that $U_{\rm s}[\varepsilon_0]=\theta^2U_{\rm B}$. Since $U_{\rm s}[\varepsilon_0]=(\theta_{\rm KN}/\theta)^{1/3}U_{\rm B}\lesssim\theta^2U_{\rm B}$ for $\gamma=\gamma_{\ast}$, and $U_{\rm s}[\varepsilon_0]=U_{\rm B}\gtrsim\theta^2U_{\rm B}$ for $\gamma=\gamma_{\rm KN}$, we have $\gamma_{\rm KN}\lesssim\gamma_{\rm b}\lesssim\gamma_{\ast}$. Below we show that $\gamma_{\rm b}=(\theta_{\rm KN}/\theta^5)\sigma_{\rm e}$.

The synchrotron spectrum is easily determined when $\gamma\lesssim\gamma_{\rm KN}$, and when $\gamma\gtrsim\gamma_{\rm b}$. When $\gamma \lesssim\gamma_{\rm KN}$, IC scattering occurs in the Thomson regime. Since $U_{\rm s}[\varepsilon_{\rm s,pk}]\simeq U_{\rm B}$, Eq. \eqref{eq:usyncT} gives
\begin{equation}
U_{\rm s}\left[\varepsilon_{\rm s}\right] = \theta^2 \left(\frac{\gamma}{\sigma_{\rm e}}\right) U_{\rm B} \propto \epsilon_{\rm s}^{1/2} \;.
\end{equation}
When $\gamma\gtrsim\gamma_{\rm b}$, IC scattering occurs in the Klein-Nishina regime. However, synchrotron is the dominant cooling channel since $U_{\rm s}[\varepsilon_0]\lesssim\theta^2U_{\rm B}$. Then Eq. \eqref{eq:usyncKN} gives
\begin{equation}
\label{eq:u1large}
U_{\rm s}\left[\varepsilon_{\rm s}\right] = \left(\frac{\gamma}{\sigma_{\rm e}}\right)U_{\rm B} \propto\varepsilon_{\rm s}^{1/2} \;.
\end{equation}
When $\gamma_{\rm KN}\lesssim\gamma\lesssim\gamma_{\rm b}$, IC scattering occurs in the Klein-Nishina regime, and IC is the dominant cooling channel since $U_{\rm s}[\varepsilon_0]\gtrsim\theta^2U_{\rm B}$. Then Eq. \eqref{eq:usyncKN} gives $U_{\rm s}[\varepsilon_{\rm s}] = (\theta^2U_{\rm B}/U_{\rm s}[\varepsilon_0])(\gamma/\sigma_{\rm e})U_{\rm B}$, which can be easily calculated once $U_{\rm s}[\varepsilon_0]$ is known. Since $\gamma\lesssim\gamma_{\rm b}\lesssim\gamma_{\ast}$, we have $\gamma_0[\gamma]\gtrsim\gamma_{\ast}\gtrsim\gamma_{\rm b}$. Then Eq. \eqref{eq:u1large} gives $U_{\rm s}[\varepsilon_0]=(\gamma_0/\sigma_{\rm e})U_{\rm B}=(\theta_{\rm KN}/\theta)^{1/2}(\sigma_{\rm e}/\gamma)^{1/2}U_{\rm B}$. Then
\begin{equation}
\label{eq:u1small}
U_{\rm s}\left[\varepsilon_{\rm s}\right] = \left(\frac{\theta^5}{\theta_{\rm KN}}\right)^{1/2}\left(\frac{\gamma}{\sigma_{\rm e}}\right)^{3/2}U_{\rm B}\propto \epsilon_{\rm s}^{3/4} \;.
\end{equation}
Eqs. \eqref{eq:u1large} and \eqref{eq:u1small} should match at $\gamma_{\rm b}$, which gives
\begin{equation}
\gamma_{\rm b} = \left(\frac{\theta_{\rm KN}}{\theta^5}\right)\sigma_{\rm e}\;.
\end{equation}
One can easily verify that $U_{\rm s}[\varepsilon_0]=\theta^2U_{\rm B}$ for $\gamma=\gamma_{\rm b}$. Then particle cooling is dominated by inverse Compton for Lorentz factors $\gamma\lesssim\gamma_{\rm b}$, and by synchrotron for $\gamma\gtrsim\gamma_{\rm b}$.

The IC spectrum has a low energy break when IC scattering transitions from the Thomson regime (for $\gamma\lesssim\gamma_{\rm KN}$) to the Klein-Nishina regime (for $\gamma\gtrsim\gamma_{\rm KN}$). Another break occurs when cooling transitions from the IC dominated regime (for $\gamma\lesssim\gamma_{\rm b}$) to the synchrotron dominated regime (for $\gamma\gtrsim\gamma_{\rm b}$). In the synchrotron dominated regime (for $\gamma\gtrsim\gamma_{\rm b}$), additional breaks occur when the Klein-Nishina threshold energy, $\varepsilon_0=m_{\rm e}c^2/\gamma$, passes through a break of the synchrotron spectrum.

When $\gamma\lesssim\gamma_{\rm b}$, cooling is dominated by IC, and Eq. \eqref{eq:uICKN} immediately gives
\begin{equation}
U_{\rm IC}\left[\varepsilon_{\rm IC}\right] =\left(\frac{\gamma}{\sigma_{\rm e}}\right)U_{\rm B}\;.
\end{equation}
Then $U_{\rm IC}\propto\gamma\propto\varepsilon_{\rm IC}^{1/2}$ for $\gamma\lesssim\gamma_{\rm KN}$, and $U_{\rm IC}\propto\varepsilon_{\rm IC}$ for $\gamma_{\rm KN}\lesssim\gamma\lesssim\gamma_{\rm b}$. When $\gamma_{\rm b}\lesssim\gamma\lesssim\sigma_{\rm e}$, cooling is dominated by synchrotron, and IC scattering occurs in the Klein-Nishina regime. Since $U_{\rm s}[\varepsilon_0]\lesssim\theta^2 U_{\rm B}$, Eq. \eqref{eq:uICKN} gives $U_{\rm IC}[\varepsilon_{\rm IC}]=(U_{\rm s}[\varepsilon_0]/\theta^2U_{\rm B})(\gamma/\sigma_{\rm e})U_{\rm B}$, which can be easily calculated once $U_{\rm s}[\varepsilon_0]$ is known. There are two cases: (i) if $\gamma_{\rm b}\lesssim\gamma\lesssim (\theta^9/\theta_{\rm KN})\sigma_{\rm e}$, we have $\gamma_{\rm b}\lesssim\gamma_0[\gamma]\lesssim\sigma_{\rm e}$. Then Eq. \eqref{eq:u1large} gives $U_{\rm s}[\varepsilon_0]=(\gamma_0/\sigma_{\rm e})U_{\rm B} = (\theta_{\rm KN}/\theta)^{1/2}(\sigma_{\rm e}/\gamma)^{1/2} U_{\rm B}$, and
\begin{equation}
U_{\rm IC}\left[\varepsilon_{\rm IC}\right]=\left(\frac{\theta_{\rm KN}}{\theta^5}\right)^{1/2}\left(\frac{\gamma}{\sigma_{\rm e}}\right)^{1/2}U_{\rm B}\;.
\end{equation}
Then $U_{\rm IC}\propto\gamma^{1/2}\propto\varepsilon_{\rm IC}^{1/2}$. On the other hand, (ii) if $\gamma\gtrsim (\theta^9/\theta_{\rm KN})\sigma_{\rm e}$, we have $\gamma_{\rm KN}\lesssim\gamma_0[\gamma]\lesssim\gamma_{\rm b}$. Then Eq. \eqref{eq:u1small} gives $U_{\rm s}[\varepsilon_0] =(\theta^5/\theta_{\rm KN})^{1/2}(\gamma_0/\sigma_{\rm e})^{3/2}U_{\rm B}= (\theta^7\theta_{\rm KN})^{1/4}(\sigma_{\rm e}/\gamma)^{3/4}U_{\rm B}$, and
\begin{equation}
U_{\rm IC}\left[\varepsilon_{\rm IC}\right]=\left(\frac{\theta_{\rm KN}}{\theta}\right)^{1/4}\left(\frac{\gamma}{\sigma_{\rm e}}\right)^{1/4}U_{\rm B}\;.
\end{equation}
Then $U_{\rm IC}\propto\gamma^{1/4}\propto\varepsilon_{\rm IC}^{1/4}$. Our results are summarised in Tables \ref{tab:sync5}-\ref{tab:sync6}.

Our results significantly simplify in the standard case of an isotropic pitch angle distribution, i.e. $\theta\sim 1$. The synchrotron spectrum is $U_{\rm s}=(\gamma/\sigma_{\rm e})U_{\rm B}$. The IC spectrum is $U_{\rm IC}=(\gamma/\sigma_{\rm e})U_{\rm B}$ for $\gamma\lesssim \gamma_{\rm KN}$, and $U_{\rm IC}=\theta_{\rm KN}^{1/2}(\gamma/\sigma_{\rm e})^{1/2}U_{\rm B}$ for $ \gamma_{\rm KN}\lesssim\gamma\lesssim\sigma_{\rm e}$. Then one recovers the familiar result that $U_{\rm s}\propto\varepsilon_{\rm s}^{1/2}$, and $U_{\rm IC}\propto\varepsilon_{\rm IC}^{1/2}$.

\subsubsection{Small pitch angles}
\label{sec:pitchKNsmall}

Next we consider the regime of small pitch angles,
\begin{equation}
\theta_{\rm KN}\lesssim\theta\lesssim \theta_{\rm KN}^{1/3}\;.
\end{equation}
As we show in the following, in this regime IC dominates the cooling for all the particles. Then synchrotron is radiatively inefficient, i.e. $U_{\rm s}[\varepsilon_{\rm s,pk}]\lesssim U_{\rm B}$.

The synchrotron spectrum has one break. The break occurs when IC cooling transitions from the Thomson regime (for $\gamma\lesssim\gamma_{\rm KN}$) to the Klein-Nishina regime (for $\gamma\gtrsim\gamma_{\rm KN}$).

When $\gamma\gtrsim\gamma_{\rm KN}$, IC scattering occurs in the Klein-Nishina regime, and $U_{\rm s}[\varepsilon_0]\gtrsim\theta^2U_{\rm B}$. Then Eq. \eqref{eq:usyncKN} gives
\begin{equation}
\label{eq:uapprox}
\frac{U_{\rm s}\left[\varepsilon_{\rm s}\right]}{U_{\rm B}} = \frac{\theta^2\gamma}{\sigma_{\rm e}} \frac{U_{\rm B}}{U_{\rm s}\left[\varepsilon_0\right]}\;.
\end{equation}
For $\gamma_{\rm KN}\lesssim\gamma\lesssim\sigma_{\rm e}$, we have $\gamma_{\rm KN}\lesssim\gamma_0[\gamma]\lesssim\sigma_{\rm e}$. Then Eq. \eqref{eq:uapprox} has a power law solution, $U_{\rm s}[\varepsilon_{\rm s}]\propto\varepsilon_{\rm s}^\alpha\propto\gamma^{2\alpha}$, and $U_{\rm s}[\varepsilon_0]\propto\varepsilon_0^\alpha\propto\gamma^{-\alpha}$. Then $\gamma^{2\alpha}\propto\gamma^{1+\alpha}$, and therefore $\alpha=1$. The normalisation of the spectrum can be determined from Eq. \eqref{eq:uast}, which gives $U_{\rm s}[\varepsilon_{\ast}] = (\theta^5\theta_{\rm KN})^{1/6}U_{\rm B}$ when $\theta\lesssim\theta_{\rm KN}^{1/7}$. Then
\begin{equation}
\label{eq:usync1}
U_{\rm s}\left[\varepsilon_{\rm s}\right] = \left(\frac{\theta^3}{\theta_{\rm KN}}\right)^{1/2} \left(\frac{\gamma}{\sigma_{\rm e}}\right)^2 U_{\rm B} \propto\varepsilon_{\rm s} \;.
\end{equation}
Substituting $\gamma=\sigma_{\rm e}$ into Eq. \eqref{eq:usync1}, we see that that synchrotron is radiatively inefficient when $\theta\lesssim \theta_{\rm KN}^{1/3}$.

When $\gamma\lesssim\gamma_{\rm KN}$, IC scattering occurs in the Thomson regime, and $U_{\rm s}[\varepsilon_{\rm s,pk}]=(\theta^3/\theta_{\rm KN})^{1/2}U_{\rm B}$. Then Eq. \eqref{eq:usyncT} gives
\begin{equation}
U_{\rm s}\left[\varepsilon_{\rm s}\right] = \left(\theta\theta_{\rm KN}\right)^{1/2} \left(\frac{\gamma}{\sigma_{\rm e}}\right) U_{\rm B}\propto\varepsilon_{\rm s}^{1/2}\;.
\end{equation}
Since cooling is dominated by IC, Eq. \eqref{eq:uICKN} immediately gives
\begin{equation}
U_{\rm IC}\left[\varepsilon_{\rm IC}\right]=\left(\frac{\gamma}{\sigma_{\rm e}}\right)U_{\rm B}\;.
\end{equation}
Then $U_{\rm IC}\propto\gamma\propto\varepsilon_{\rm IC}^{1/2}$ for $\gamma\lesssim\gamma_{\rm KN}$, and $U_{\rm IC}\propto\varepsilon_{\rm IC}$ for $\gamma_{\rm KN}\lesssim\gamma\lesssim\sigma_{\rm e}$. Our results are summarised in Table \ref{tab:sync2}.

\subsubsection{Intermediate pitch angles}
\label{sec:pitchKNint}

Finally we consider the regime of intermediate pitch angles,
\begin{equation}
\theta_{\rm KN}^{1/3}\lesssim\theta\lesssim \theta_{\rm KN}^{1/7}\;.
\end{equation}
In this regime synchrotron dominates the cooling of the most energetic particles, i.e. $U_{\rm s}[\varepsilon_0]\lesssim\theta^2U_{\rm B}$ for $\gamma=\sigma_{\rm e}$. Substituting $\gamma=\sigma_{\rm e}$ into Eq. \eqref{eq:usyncKN} gives $U_{\rm s}[\varepsilon_{\rm s,pk}]=U_{\rm B}$. However, IC dominates the cooling of particles with $\gamma=\gamma_{\ast}$, i.e. $U_{\rm s}[\varepsilon_0]\gtrsim\theta^2U_{\rm B}$ for $\gamma=\gamma_{\ast}$. Then radiation has a different spectrum with respect to the case of large pitch angles, i.e. $\theta_{\rm KN}^{1/7}\lesssim\theta\lesssim 1$.

The synchrotron spectrum has three breaks. A low energy break occurs when IC cooling transitions from the Thomson regime (for $\gamma\lesssim\gamma_{\rm KN}$) to the Klein-Nishina regime (for $\gamma\gtrsim\gamma_{\rm KN}$). A high energy break occurs when cooling transitions from the IC dominated regime (for $\gamma\lesssim\gamma_{\rm b}$) to the synchrotron dominated regime (for $\gamma\gtrsim\gamma_{\rm b}$). In this regime of pitch angles, we have $\gamma_{\rm b}\gtrsim\gamma_{\ast}$. An intermediate energy break appears at $\gamma=\gamma_{\rm i}$, when the Klein-Nishina threshold energy, $\varepsilon_0[\gamma]=m_{\rm e}c^2/\gamma$, passes through the high energy spectral break, $\varepsilon_{\rm s}[\gamma_{\rm b}]$. Then $\varepsilon_0[\gamma_{\rm i}]=\varepsilon_{\rm s}[\gamma_{\rm b}]$, which gives $\gamma_{\rm b}=\gamma_0[\gamma_{\rm i}]$. In the following we show that $\gamma_{\rm b}=(\theta_{\rm KN}/\theta^3)^{1/2}\sigma_{\rm e}$, and $\gamma_{\rm i}=\theta^2\sigma_{\rm e}$.\footnote{When the pitch angle is $\theta=\theta_{\rm KN}^{1/7}$, we have $\gamma_{\rm b}=\gamma_{\ast}$. The high energy break merges with the intermediate energy break, i.e. $\gamma_{\rm b}=\gamma_{\rm i}=\gamma_{\ast}$. For larger pitch angles, $\theta\gtrsim\theta_{\rm KN}^{1/7}$, we have only one break at $\gamma_{\rm b}\lesssim\gamma_{\ast}$.}

The synchrotron spectrum is easily determined when $\gamma\lesssim\gamma_{\rm KN}$, and when $\gamma\gtrsim\gamma_{\rm b}$. When $\gamma \lesssim\gamma_{\rm KN}$, IC scattering occurs in the Thomson regime, and $U_{\rm s}[\varepsilon_{\rm s,pk}]=U_{\rm B}$. Then Eq. \eqref{eq:usyncT} gives
\begin{equation}
U_{\rm s}\left[\varepsilon_{\rm s}\right] = \theta^2 \left(\frac{\gamma}{\sigma_{\rm e}}\right) U_{\rm B}\propto\varepsilon_{\rm s}^{1/2}\;.
\end{equation}
When $\gamma\gtrsim\gamma_{\rm b}$, IC scattering occurs in the Klein-Nishina regime, and $U_{\rm s}[\varepsilon_0]\lesssim\theta^2U_{\rm B}$. Then Eq. \eqref{eq:usyncKN} gives
\begin{equation}
\label{eq:uint1}
U_{\rm s}\left[\varepsilon_{\rm s}\right] = \left(\frac{\gamma}{\sigma_{\rm e}}\right)U_{\rm B}\propto\varepsilon_{\rm s}^{1/2}\;.
\end{equation}
When $\gamma_{\rm KN}\lesssim\gamma\lesssim\gamma_{\rm i}$, IC scattering occurs in the Klein-Nishina regime, and $U_{\rm s}[\varepsilon_0]\gtrsim\theta^2U_{\rm B}$. Then Eq. \eqref{eq:usyncKN} gives $U_{\rm s}[\varepsilon_{\rm s}] = (\theta^2U_{\rm B}/U_{\rm s}[\varepsilon_0])(\gamma/\sigma_{\rm e})U_{\rm B}$, which can be easily calculated once $U_{\rm s}[\varepsilon_0]$ is known. Since $\gamma\lesssim\gamma_{\rm i}$, and $\gamma_0[\gamma_{\rm i}]=\gamma_{\rm b}$, we have $\gamma_0[\gamma]\gtrsim\gamma_{\rm b}$. Then Eq. \eqref{eq:uint1} gives $U_{\rm s}[\varepsilon_0]=(\gamma_0/\sigma_{\rm e})U_{\rm B}=(\theta_{\rm KN}/\theta)^{1/2}(\sigma_{\rm e}/\gamma)^{1/2}U_{\rm B}$. Then
\begin{equation}
\label{eq:uint2}
U_{\rm s}\left[\varepsilon_{\rm s}\right] = \left(\frac{\theta^5}{\theta_{\rm KN}}\right)^{1/2}\left(\frac{\gamma}{\sigma_{\rm e}}\right)^{3/2}U_{\rm B}\propto\varepsilon_{\rm s}^{3/4}\;.
\end{equation}
When $\gamma_{\rm i}\lesssim\gamma\lesssim\gamma_{\rm b}$, we have $\gamma_{\rm i}\lesssim\gamma_0[\gamma]\lesssim\gamma_{\rm b}$. The same arguments used to derive Eq. \eqref{eq:usync1} give
\begin{equation}
\label{eq:uint3}
U_{\rm s}\left[\varepsilon_{\rm s}\right] = \left(\frac{\theta^3}{\theta_{\rm KN}}\right)^{1/2} \left(\frac{\gamma}{\sigma_{\rm e}}\right)^2 U_{\rm B}\propto\varepsilon_{\rm s}\;.
\end{equation}
The Lorentz factors $\gamma_{\rm b}$ and $\gamma_{\rm i}$ can be determined by requiring that $U_{\rm s}$ is a continuous function of $\gamma$. Then
\begin{equation}
\gamma_{\rm b}= \left(\frac{\theta_{\rm KN}}{\theta^3}\right)^{1/2} \sigma_{\rm e}
\end{equation}
and
\begin{equation}
\gamma_{\rm i}= \theta^2\sigma_{\rm e} \;.
\end{equation}
The IC spectrum is easily determined when $\gamma\lesssim\gamma_{\rm b}$. Since cooling is dominated by IC, Eq. \eqref{eq:uICKN} immediately gives
\begin{equation}
U_{\rm IC}\left[\varepsilon_{\rm IC}\right] =\left(\frac{\gamma}{\sigma_{\rm e}}\right)U_{\rm B}\;.
\end{equation}
Then $U_{\rm IC}\propto\gamma\propto\varepsilon_{\rm IC}^{1/2}$ for $\gamma\lesssim\gamma_{\rm KN}$, and $U_{\rm IC}\propto\varepsilon_{\rm IC}$ for $\gamma_{\rm KN}\lesssim\gamma\lesssim\gamma_{\rm b}$. When $\gamma_{\rm b}\lesssim\gamma\lesssim\sigma_{\rm e}$, cooling is dominated by synchrotron, and IC scattering occurs in the Klein-Nishina regime. Since $U_{\rm s}[\varepsilon_0]\lesssim\theta^2 U_{\rm B}$, Eq. \eqref{eq:uICKN} gives $U_{\rm IC}[\varepsilon_{\rm IC}]=(U_{\rm s}[\varepsilon_0]/\theta^2U_{\rm B})(\gamma/\sigma_{\rm e})U_{\rm B}$, which can be easily calculated once $U_{\rm s}[\varepsilon_0]$ is known. There are two cases: (i) if $\gamma_{\rm b} \lesssim \gamma\lesssim (\theta_{\rm KN}/\theta^5)\sigma_{\rm e}$, we have $\gamma_{\rm i}\lesssim\gamma_0[\gamma]\lesssim\gamma_{\rm b}$. Then Eq. \eqref{eq:uint3} gives $U_{\rm s}[\varepsilon_0]=(\theta^3/\theta_{\rm KN})^{1/2} (\gamma_0/\sigma_{\rm e})^2 U_{\rm B} = (\theta\theta_{\rm KN})^{1/2}(\sigma_{\rm e}/\gamma)U_{\rm B}$, and
\begin{equation}
U_{\rm IC}\left[\varepsilon_{\rm IC}\right] = \left(\frac{\theta_{\rm KN}}{\theta^3}\right)^{1/2}U_{\rm B}\;.
\end{equation}
Then $U_{\rm IC}\propto\gamma^0\propto\varepsilon_{\rm IC}^0$. On the other hand, (ii) if $\gamma\gtrsim (\theta_{\rm KN}/\theta^5)\sigma_{\rm e}$, we have $\gamma_{\rm KN}\lesssim\gamma_0[\gamma]\lesssim\gamma_{\rm i}$. Then Eq. \eqref{eq:uint2} gives $U_{\rm s}[\varepsilon_0] =(\theta^5/\theta_{\rm KN})^{1/2}(\gamma_0/\sigma_{\rm e})^{3/2}U_{\rm B} = (\theta^7\theta_{\rm KN})^{1/4}(\sigma_{\rm e}/\gamma)^{3/4}U_{\rm B}$, and
\begin{equation}
U_{\rm IC}\left[\varepsilon_{\rm IC}\right] = \left(\frac{\theta_{\rm KN}}{\theta}\right)^{1/4}\left(\frac{\gamma}{\sigma_{\rm e}}\right)^{1/4}U_{\rm B} \;.
\end{equation}
Then $U_{\rm IC}\propto\gamma^{1/4}\propto\varepsilon_{\rm IC}^{1/4}$. Our results are summarised in Tables \ref{tab:sync3}-\ref{tab:sync4}.

\section{Astrophysical implications}
\label{sec:astro}

We now apply our results to the modelling of blazars and GRBs. In Section \ref{sec:KN} we neglected factors $\sim 1$, and below for numerical estimates we will use better approximate coefficients in $P_{\rm s}$, $P_{\rm IC}$, $\varepsilon_{\rm s}$, and $\varepsilon_{\rm IC}$:
\begin{align}
P_{\rm s} & \simeq 2c\sigma_{\rm T}\theta^2U_{\rm B}\gamma^2 \\
P_{\rm IC} & \simeq \frac{4}{3}c\sigma_{\rm T} U_{\rm s,av}\gamma^2 \\
\varepsilon_{\rm s} & \simeq \frac{1}{2}\theta\gamma^2\left(\frac{B}{B_{\rm q}}\right)m_{\rm e}c^2 \\ 
\varepsilon_{\rm IC} & \simeq \max\left[\frac{4}{3}\gamma^2 \varepsilon_{\rm s,pk}, \frac{1}{2}\gamma m_{\rm e}c^2\right]  \;.
\end{align}
Then  $\theta_{\rm KN}$ is changed from Eq. \eqref{eq:alphaKN} by a factor of $3/4$: $\theta_{\rm KN}=(3/4)(1/\sigma_{\rm e}^3)(B_{\rm q}/B)$.

\subsection{Blazars}

Blazar spectra are characterised by two broad non-thermal components, the first one peaking at IR-optical-UV frequencies, and the second one peaking in the gamma-rays. Spectra follow a well known sequence, with fainter objects peaking at higher frequencies \citep[e.g.][]{Fossati1998, Ghisellini+2017}. We focus on the faintest blazars in the sequence, i.e. BL Lac objects, where the two spectral components are likely emitted by the same population of non-thermal electrons via synchrotron-self-Compton \citep[e.g.][]{Maraschi1992, Tavecchio1998, Tavecchio2010}.\footnote{The two spectral components are emitted by the same electrons also in the brightest blazars, i.e. Flat Spectrum Radio Quasars (FSRQ). However, gamma-rays in FSRQ are likely produced by IC scattering off an external photon field \citep[e.g.][]{Sikora1994, Sikora2009, GhiselliniTavecchio2009}.}

\begin{figure}
\centering
\includegraphics[width=0.47\textwidth]{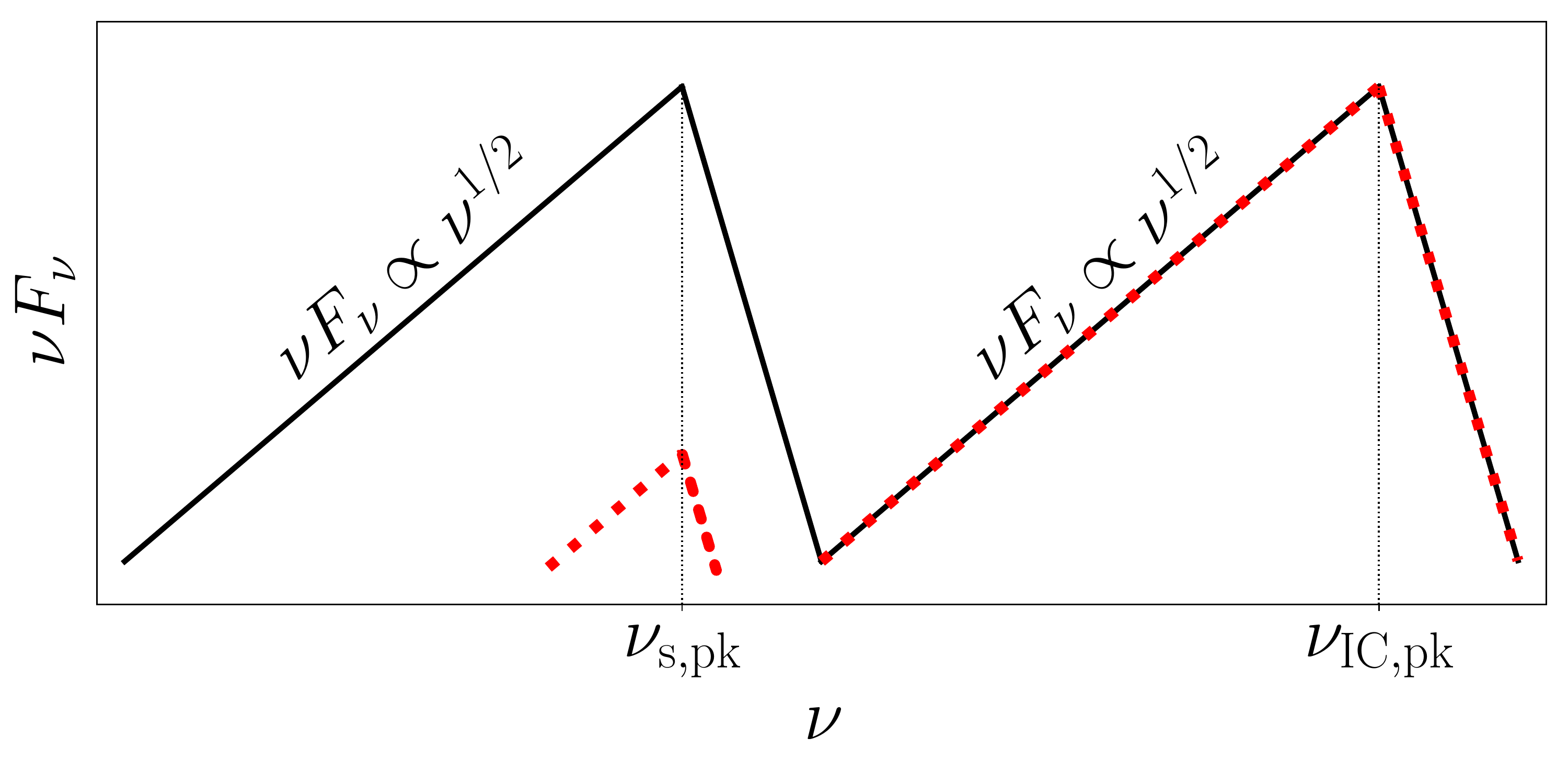}
\caption{Sketch of the synchrotron-self-Compton spectrum of BL Lacs (see also Table \ref{tab:sync1}). Solid line: particle pitch angles $\theta\gtrsim 0.1$, producing the typical emission, with comparable synchrotron and IC luminosities. Dotted line: particle pitch angles $\theta\ll 0.1$, producing orphan gamma-ray flares. The ratio between the synchrotron and IC luminosities is $L_{\rm s}/L_{\rm IC}\sim\theta$ (see Eq. \ref{eq:LL}). The ratio between the peak frequencies is $\nu_{\rm IC, pk}/\nu_{\rm s, pk}\sim\sigma_{\rm e}^2$. In the simple model we consider (i.e., $\delta$-function injection) the spectrum would cut off exponentially above the peak.}
\label{fig:AGN}
\end{figure}

We argue that synchrotron-self-Compton emission from a population of fast cooling electrons in a magnetically-dominated plasma can naturally explain the common features of typical BL Lac spectra \citep[for a large compilation of BL Lac spectra, see e.g.][]{Tavecchio2010}. First, at frequencies below the peak both synchrotron and IC spectra are well described by a power law, $\nu F_\nu\propto\nu^\alpha$, with a soft spectral slope $\alpha\sim 1/2$. Such a slope is naturally produced by a population of fast cooling electrons when Klein-Nishina effects are minor. Second, the luminosities of the UV and gamma-ray peaks are comparable (typically within an order of magnitude). In fast cooling magnetically-dominated plasmas, the magnetic energy is converted into synchrotron radiation on the light crossing time of the system. Since the radiation escape time is equal to the dissipation time, the radiation energy density is equal to the magnetic energy density. If the pitch angle is not too small (see Eq. \eqref{eq:LL} below), particles radiate a comparable amount of energy via synchrotron and IC.\footnote{Alternatively, comparable UV and gamma-ray luminosities may be produced also in weakly magnetised plasmas, i.e. in the regime $U_{\rm e}\gg U_{\rm B}$. Producing comparable luminosities requires that $U_{\rm s}\sim U_{\rm B}$. Then particles should radiate only a small fraction $\sim U_{\rm B}/U_{\rm e}$ of their energy. This requires an undesirable fine tuning of the cooling time, i.e. $t_{\rm cool}\sim (U_{\rm e}/U_{\rm B}) t_{\rm dyn}$.} A similar explanation for the common features of BL Lac spectra has been discussed by \citet[][]{SobacchiLyubarsky2020}. The spectrum is sketched in Figure \ref{fig:AGN}.

Two basic observed properties of synchrotron-self-Compton emission of blazars are (i) the ratio between the IC and the synchrotron peak energies, $\zeta=E_{\rm IC,pk}/E_{\rm s,pk}\sim 10^9$, and (ii) the isotropic equivalent total luminosity, $L_{\rm iso}=L_{\rm s}+L_{\rm IC}\sim 10^{45}{\rm\; erg\; s}^{-1}$. The quoted values are meant to represent a ``typical'' BL Lac \citep[e.g.][]{Tavecchio2010}. We also normalise the bulk Lorentz factor of the emitting plasma to a typical value of $\Gamma\sim 10$ \citep[e.g.][]{Hovatta2009, Lister2009}. We consider dissipation radii $R\gtrsim 10^{16}{\rm\; cm}$, consistent with a variability timescale of the light curve $t_{\rm var}\sim R/2c\Gamma^2\sim 2\times 10^3 R_{16}\Gamma_{10}^{-2}{\rm\; s}$. Hereafter we use the notation $\zeta_9\equiv\zeta/10^9$, $L_{45}\equiv L_{\rm iso}/10^{45}{\rm\; erg\; s}^{-1}$, $R_{16}\equiv R/10^{16}{\rm\; cm}$, and $\Gamma_{10}\equiv\Gamma/10$.

Below we describe the parameters of our model that would give the observed blazar spectra. Since in the fast cooling regime the dissipated magnetic energy $\sim U_{\rm B}$ is promptly converted into radiation, the total luminosity is $L_{\rm iso}\sim c\Gamma^2B^2R^2$. Then the magnetic field in the rest frame of the plasma is
\begin{equation}
\label{eq:BAGN}
B \sim 2 \; L_{45}^{1/2} \Gamma_{10}^{-1} R_{16}^{-1} {\rm\; G}\;.
\end{equation}
Soft blazar spectra may be produced when the electrons are cooling due to IC scattering in the Thomson cooling regime, with $\theta \lesssim \theta_{\rm KN}$. The properties of the emitted radiation are summarised in Table \ref{tab:sync1}. Since $E_{\rm IC,pk}/E_{\rm s,pk}\simeq (4/3)\sigma_{\rm e}^2$, we find that 
\begin{equation}
\sigma_{\rm e} \sim 3\times 10^4 \zeta_9^{1/2}
\end{equation}
and
\begin{equation}
\label{eq:alphaKNBL}
\theta_{\rm KN} \sim 0.9\; L_{45}^{-1/2} \Gamma_{10} R_{16} \zeta_9^{-3/2} \;.
\end{equation}
Then the condition that $\theta\lesssim\theta_{\rm KN}$ may be satisfied even for large pitch angles. Note that in electron-proton plasmas the overall magnetisation is $\sigma=(m_{\rm e}/m_{\rm p})\sigma_{\rm e}\sim 10$.

The cooling timescale for electrons with Lorentz factor $\gamma=\sigma_{\rm e}$ is given by $t_{\rm cool}^{\rm pk}/t_{\rm dyn}=(1/2)\theta^{-1}\sigma_{\rm e}^{-1}\ell_{\rm B}^{-1}$, where $\ell_{\rm B}=\sigma_{\rm T}U_{\rm B}t_{\rm dyn}/m_{\rm e}c$. Since the dynamical time in the rest frame of the plasma is $t_{\rm dyn}=R/\Gamma c$, we have
\begin{equation}
\ell_{\rm B} \sim 10^{-4}L_{45}\Gamma_{10}^{-3}R_{16}^{-1} \;.
\end{equation}
Then
\begin{equation}
\label{eq:tratioAGN}
\frac{t_{\rm cool}^{\rm pk}}{t_{\rm dyn}} \sim 0.2\; L_{45}^{-1} \Gamma_{10}^3 R_{16} \zeta_9^{-1/2} \theta^{-1} \;.
\end{equation}
Note that the ratio of magnetic and electron energy densities is $U_{\rm B}/U_{\rm e}\sim t_{\rm dyn}/t_{\rm cool}^{\rm pk}$ (see Eq. \ref{eq:Ue}). The condition for fast cooling, $t_{\rm cool}^{\rm pk}\lesssim t_{\rm dyn}$, can be satisfied if dissipation occurs at relatively small radii, $R\sim 10^{16}{\rm\; cm}$. At these radii, the inferred $t_{\rm cool}^{\rm pk}/t_{\rm dyn}$ may be further reduced if the bulk Lorentz factor of the jet is smaller than its asymptotic value $\Gamma\sim 10$. Imaging of radio emission from extragalactic jets suggests that the bulk acceleration may be still in progress on sub-parsec scales \citep[e.g.][]{Boccardi+2016, Mertens+2016}.

The peak energy of the observed synchrotron radiation is $E_{\rm s,pk}=\Gamma\varepsilon_{\rm s,pk}\simeq (1/2)\Gamma\theta\sigma_{\rm e}^2(B/B_{\rm q})m_{\rm e}c^2$, which gives
\begin{equation}
E_{\rm s,pk} \sim 80\; L_{45}^{1/2} R_{16}^{-1} \zeta_9 \theta {\rm\; eV} \;.
\end{equation}
The peak energy of the observed IC radiation is $E_{\rm IC,pk}=\Gamma\varepsilon_{\rm IC,pk}\simeq (1/2) \Gamma\theta\sigma_{\rm e}^4(B/B_{\rm q})m_{\rm e}c^2$, which gives
\begin{equation}
E_{\rm IC,pk} \sim 80\; L_{45}^{1/2} R_{16}^{-1} \zeta_9^2 \theta {\rm\; GeV} \;.
\end{equation}
Then the synchrotron radiation peaks in the UV, and the IC radiation peaks in the gamma-rays, as observed.

The ratio between the synchrotron luminosity and the IC luminosity is
\begin{equation}
\label{eq:LL}
\frac{L_{\rm s}}{L_{\rm IC}} \sim \theta \;.
\end{equation}
The typical BL Lac spectra are characterised by comparable UV and gamma-ray luminosities (within a factor of ten). This naturally occurs if the emitting particles have a nearly isotropic pitch angle distribution, i.e. $\theta\gtrsim 0.1$. The effects that control the pitch angle distribution are discussed in Section \ref{sec:orphan}.

Fitting the spectra of individual BL Lacs under the assumption of isotropic particles, one typically infers a low ratio of the magnetic and electron energy densities, $U_{\rm B}/U_{\rm e}\sim 0.01$ \citep[e.g.][]{TavecchioGhisellini2016}. Since the synchrotron frequency and power depend on the component of the magnetic field perpendicular to the particle velocity, $B\sin\theta$, this result is very sensitive to the anisotropy of the emitting particles. For pitch angles $\theta\sim 0.1$, the inferred value of $U_{\rm B}/U_{\rm e}$ would increase by a factor of $\theta^{-2}\sim 100$, becoming of order unity. In turn, for $\theta\sim 0.1$ our model gives $U_{\rm B}/U_{\rm e}\sim t_{\rm dyn}/t_{\rm cool}^{\rm pk}\sim 1$ (see Eq. \ref{eq:tratioAGN}) and $L_{\rm s}/L_{\rm IC}\sim 0.1$ (see Eq. \ref{eq:LL}). Then pitch angles $\theta\sim 0.1$ may be consistent with observational constraints.

In our discussion, we have neglected Klein-Nishina effects on IC scattering. Since  in the Klein-Nishina regime the IC power is suppressed, one finds that $L_{\rm s}>\theta L_{\rm IC}$. Then the synchrotron and IC luminosities may be comparable (within a factor of ten) even for pitch angles $\theta< 0.1$. IC scattering occurs deep into the Klein-Nishina regime in the so-called hard-TeV BL Lacs \citep[e.g.][]{Costamante+2018, Biteau+2020}. Interestingly, in these objects the ratio of magnetic and electron energy densities inferred from the spectral modelling under the assumption of isotropic particles is very low, $U_{\rm B}/U_{\rm e}\sim 10^{-3}-10^{-4}$. Testing our model on hard-TeV BL Lacs is an interesting direction for future work.

\subsubsection{Orphan gamma-ray flares}
\label{sec:orphan}

Since UV and gamma-rays are emitted by the same particles, one expects the light curves in the two bands to be correlated. This picture is challenged by the rare occurrence of orphan gamma-ray flares, i.e. flares lacking a luminous low energy counterpart \citep[e.g.][]{Krawczynski2004, Blazejowski2005}.

We argue that orphan gamma-ray flares may be associated with rare events when the emitting particles have very small pitch angles.\footnote{\citet[][]{Ghisellini+2009} also suggested that orphan gamma-ray flares are produced by particles accelerated along the magnetic field lines. These authors argued that anisotropic particle distributions are produced via magneto-centrifugal acceleration.} When $\theta\ll 0.1$, the IC luminosity is much larger than the synchrotron luminosity, i.e. $L_{\rm IC}\gg L_{\rm s}\sim\theta L_{\rm IC}$ (see Eq. \ref{eq:LL}). Then the gamma-ray emission may have a suppressed UV counterpart. For a luminous flare with $L_{\rm IC}\sim 10^{46}{\rm\; erg\; s}^{-1}$, particles with a small pitch angle $\theta\sim 0.02$ are in the fast cooling regime (see Eq. \ref{eq:tratioAGN}). The spectrum is sketched in Figure \ref{fig:AGN}. A detailed study of orphan gamma-ray flares has been presented elsewhere \citep[][]{Sobacchi+2021}.

The pitch angle distribution of the emitting particles may be regulated by the level of magnetic field fluctuations (as compared to the mean field) from which turbulence develops. Larger initial fluctuations produce more isotropic particle distributions \citep[][]{Comisso+2020, Sobacchi+2021}. A complementary possibility, yet to be tested with first principles simulations, is that the pitch angle distribution depends on the plasma composition. In electron-proton plasmas, the pitch angle distribution may be isotropised by a kinetic instability that is absent in electron-positron plasmas \citep[][]{SobacchiLyubarsky2019}.

\subsection{Gamma-Ray Bursts}

At frequencies below the peak, the spectrum of the GRB prompt emission is well described by a power law, $\nu F_\nu\propto\nu^\alpha$, with a typical spectral slope $\alpha\sim 1$ \citep[e.g.][]{Preece+2000, Kaneko+2006, Nava+2011, Gruber+2014}. The spectral slope is significantly harder than $\alpha=1/2$, which is the slope produced by fast cooling electrons when synchrotron is the dominant cooling channel.

The typical spectral slope of the GRB prompt emission spectra can be produced by synchrotron if the emitting electrons radiate most of their energy via IC, and the scattering occurs in the Klein-Nishina regime \citep[e.g.][]{Derishev+2001, Bosnjak+2009, Nakar+2009, Daigne+2011}. If the particle pitch angle distribution is isotropic, this requires the radiation energy density to be much larger than the magnetic energy density, i.e. $U_{\rm s}\gg U_{\rm B}$ (otherwise cooling would be dominated by synchrotron, and $\alpha=1/2$). Then such a scenario is not viable in magnetically-dominated plasmas, where necessarily $U_{\rm s}\lesssim U_{\rm B}$. By contrast, if the pitch angle $\theta$ is small, the condition for the IC cooling dominance becomes $U_{\rm s} \gg \theta^2 U_{\rm B}$. This condition may be easily satisfied even in magnetically-dominated plasmas.

In the following we discuss the parameters of the emitting plasma that could give synchrotron emission with two observed properties: (i) the peak energy of the observed radiation, $E_{\rm s,pk}\sim 1{\rm\; MeV}$, and (ii) the isotropic equivalent of the GRB luminosity, $L_{\rm iso}\sim 10^{52}{\rm\; erg\; s}^{-1}$. The quoted values are meant to represent a ``typical'' GRB. We also normalise the bulk Lorentz factor of the emitting plasma to a typical value of $\Gamma\sim 300$ \citep[e.g.][]{LithwickSari2001}. We consider sufficiently large dissipation radii $R\gtrsim 10^{15}{\rm\; cm}$, outside the jet photosphere. At these radii, the expected variability timescale of the light curve is $t_{\rm var}\sim R/2c\Gamma^2\sim 0.2\;R_{15}\Gamma_{300}^{-2}{\rm\; s}$. Hereafter we use the notation $E_6\equiv E_{\rm s,pk}/1{\rm\; MeV}$, $L_{52}\equiv L_{\rm iso}/10^{52}{\rm\; erg\; s}^{-1}$, $R_{15}\equiv R/10^{15}{\rm\; cm}$, and $\Gamma_{300}\equiv\Gamma/300$.

Assuming that a large fraction of the available electromagnetic jet energy is converted into synchrotron radiation (this is expected if the pitch angle is not too small, see Eq. \eqref{eq:alphalim} below), the observed luminosity is $L_{\rm iso}\sim c\Gamma^2B^2R^2$. Then the magnetic field in the rest frame of the plasma is
\begin{equation}
B\sim 2 \; L_{52}^{1/2} \Gamma_{300}^{-1} R_{15}^{-1} {\rm\; kG}\;.
\end{equation}
The peak energy of the observed radiation is $E_{\rm s,pk}=\Gamma\varepsilon_{\rm s,pk}\simeq (1/2)\Gamma\theta\sigma_{\rm e}^2(B/B_{\rm q})m_{\rm e}c^2$, which gives
\begin{equation}
\label{eq:sigmaeGRB}
\sigma_{\rm e} \sim 2\times 10^4 L_{52}^{-1/4}R_{15}^{1/2} E_6^{1/2} \theta^{-1/2} \;.
\end{equation}
Note that in electron-proton plasmas the overall magnetisation is $\sigma= (m_{\rm e}/m_{\rm p})\sigma_{\rm e}\sim 10$.

The IC scattering regime is determined by the critical pitch angle $\theta_{\rm KN}=(3/4)(1/\sigma_{\rm e}^3)(B_{\rm q}/B)$. For the typical parameters of GRBs, we find
\begin{equation}
\label{eq:alphaKNgrb}
\frac{\theta}{\theta_{\rm KN}} \sim 300\; L_{52}^{-1/4} \Gamma_{300}^{-1} R_{15}^{1/2} E_6^{3/2} \theta^{-1/2} \;.
\end{equation}
One can see from this equation that $\theta\gg\theta_{\rm KN}$ for any $\theta\lesssim 1$. Hence, IC scattering occurs in the Klein-Nishina regime.\footnote{It is easy to see why the scattering occurs in the Klein-Nishina regime. In the rest frame of the plasma, the energy of the photons at the peak of the spectrum is $E_{\rm s,pk}/\Gamma\sim 3\;E_6\Gamma_{300}^{-1}{\rm\; keV}$. This energy is much larger than $m_{\rm e}c^2/\sigma_{\rm e}\sim 40\;L_{52}^{1/4} R_{15}^{-1/2}E_6^{-1/2} \theta^{1/2}{\rm\; eV}$.} Cooling is dominated by synchrotron if $\theta\gtrsim (2\theta_{\rm KN}/3)^{1/3}$, or
\begin{equation}
\label{eq:alphalim}
\theta\gtrsim 0.02 \; L_{52}^{1/6} \Gamma_{300}^{2/3} R_{15}^{-1/3} E_6^{-1} \;.
\end{equation}
For smaller pitch angles, synchrotron is radiatively inefficient.

We illustrate the effect of the pitch angle anisotropy on the synchrotron spectrum assuming that $(2\theta_{\rm KN}/3)^{1/3} \lesssim\theta\lesssim (4\theta_{\rm KN}/9)^{1/5}$, which is the regime described in Table \ref{tab:sync3}. This condition requires $0.02 \; L_{52}^{1/6} \Gamma_{300}^{2/3} R_{15}^{-1/3} E_6^{-1} \lesssim\theta\lesssim 0.3\; L_{52}^{1/14} \Gamma_{300}^{2/7} R_{15}^{-1/7}E_6^{-3/7}$. In this regime of pitch angles, most of the magnetic energy is converted into synchrotron radiation, and IC losses in the Klein-Nishina regime harden the synchrotron spectrum below the peak.

\begin{figure}
\centering
\includegraphics[width=0.47\textwidth]{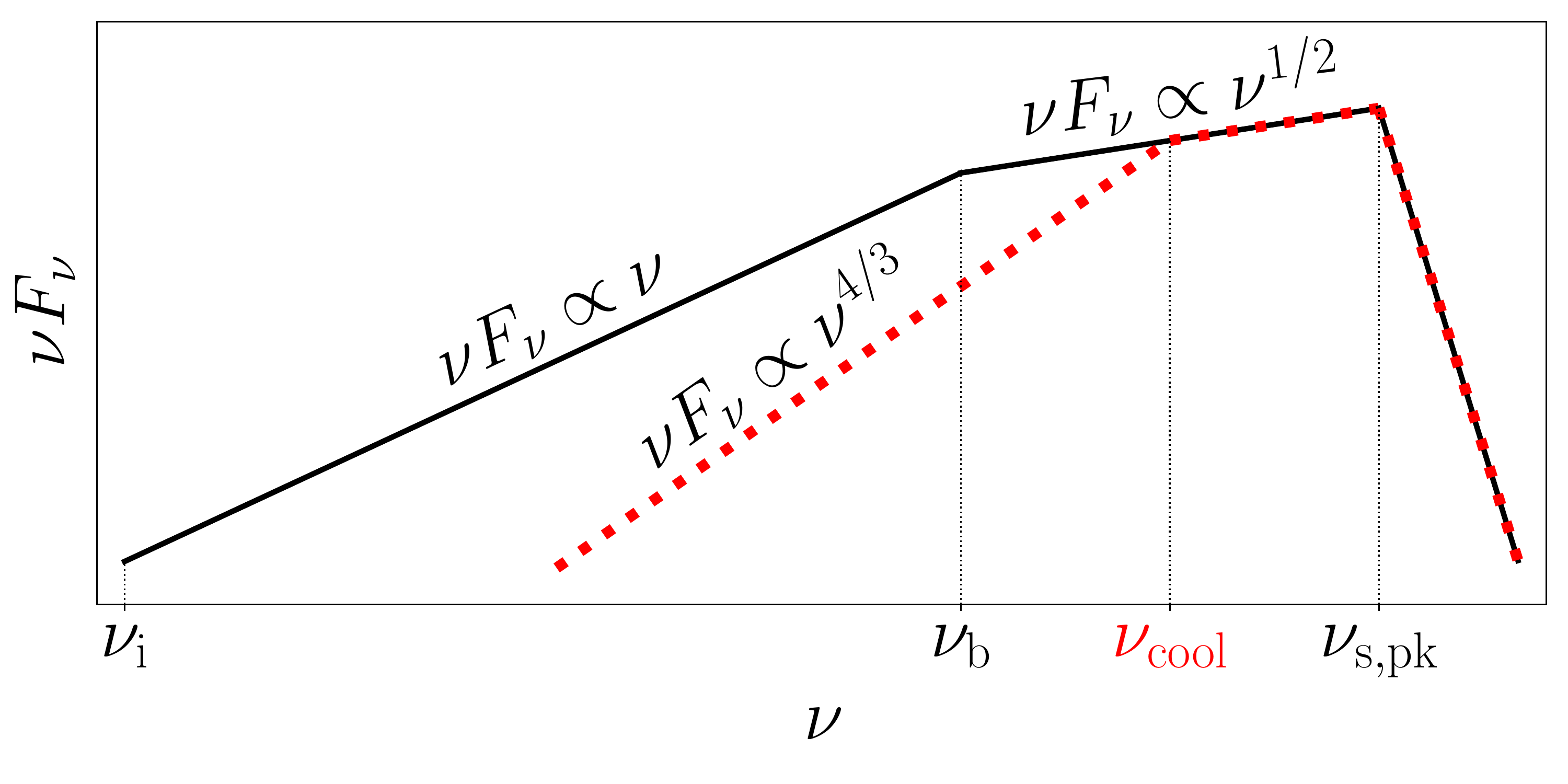}
\caption{Sketch of the synchrotron spectrum of GRB prompt emission, assuming particle pitch angles $\theta\sim 0.1$ (see also Table \ref{tab:sync3}). Solid line: dissipation radii $R\lesssim R_{\rm cool,2}$, where the cooling break occurs at a frequency $\nu_{\rm cool}\lesssim \nu_{\rm i}$. Dotted line: dissipation radii $R_{\rm cool,2}\lesssim R\lesssim R_{\rm cool,1}$, where $\nu_{\rm b}\lesssim\nu_{\rm cool}\lesssim\nu_{\rm s,pk}$. Since $\nu_{\rm s,pk}/\nu_{\rm b}\sim 15$, and $\nu_{\rm b}/\nu_{\rm i}\sim 350$ (see Eqs. \ref{eq:epsb}-\ref{eq:epsi}), the hard part of the spectrum, $\nu F_\nu\propto\nu$, extends over a broad range of frequencies. In the simple model we consider (i.e., $\delta$-function injection) the spectrum would cut off exponentially above the peak.}
\label{fig:GRB}
\end{figure}

The cooling timescale for electrons with Lorentz factor $\gamma=\sigma_{\rm e}$ is given by $t_{\rm cool}^{\rm pk}/t_{\rm dyn}=(1/2)\theta^{-2}\sigma_{\rm e}^{-1} \ell_{\rm B}^{-1}$, where $\ell_{\rm B}=\sigma_{\rm T}U_{\rm B}t_{\rm dyn}/m_{\rm e}c$. Since the dynamical time in the rest frame of the plasma is $t_{\rm dyn}=R/\Gamma c$, we have
\begin{equation}
\ell_{\rm B} \sim 0.4\; L_{52}\Gamma_{300}^{-3}R_{15}^{-1}\;.
\end{equation}
Then
\begin{equation}
\label{eq:ratio2}
\frac{t_{\rm cool}^{\rm pk}}{t_{\rm dyn}}\sim 2\times10^{-3} L_{52}^{-3/4}\Gamma_{300}^3 R_{15}^{1/2} E_6^{-1/2} \theta_{-1}^{-3/2} \;,
\end{equation}
where $\theta_{-1}\equiv\theta/0.1$. The fast cooling condition $t_{\rm cool}^{\rm pk}\lesssim t_{\rm dyn}$ is satisfied at radii of interest $R\lesssim R_{\rm cool,1}$, where
\begin{equation}
\label{eq:Rdiss2}
R_{\rm cool,1} = 2\times 10^{20} L_{52}^{3/2}\Gamma_{300}^{-6} E_6\theta_{-1}^3 {\rm\; cm} \;.
\end{equation}

The synchrotron spectrum depends on whether electrons with intermediate Lorentz factors, $(3/2)\theta^2\sigma_{\rm e}\lesssim\gamma\lesssim (2\theta_{\rm KN}/3\theta^3)^{1/2}\sigma_{\rm e}$, are fast cooling. These electrons have  $t_{\rm cool}/t_{\rm dyn}=(8\theta\theta_{\rm KN}/3)^{-1/2}\sigma_{\rm e}^{-1} \ell_{\rm B}^{-1}$, i.e.
\begin{equation}
\label{eq:ratio1}
\frac{t_{\rm cool}}{t_{\rm dyn}}\sim 9\times 10^{-3} L_{52}^{-7/8}\Gamma_{300}^{5/2} R_{15}^{3/4} E_6^{1/4}\theta_{-1}^{-3/4} \;.
\end{equation}
Note that $t_{\rm cool}/t_{\rm dyn}$ is independent of $\gamma$. The condition $t_{\rm cool}\lesssim t_{\rm dyn}$ for the intermediate $\gamma$ is stronger than $t_{\rm cool}^{\rm pk} \lesssim t_{\rm dyn}$, and it is satisfied at smaller radii $R\lesssim R_{\rm cool,2}$, where
\begin{equation}
\label{eq:Rdiss1}
R_{\rm cool,2}= 6\times 10^{17} L_{52}^{7/6}\Gamma_{300}^{-10/3}E_6^{-1/3}\theta_{-1} {\rm\; cm} \;.
\end{equation}

The expected spectrum in the two cases $R\lesssim R_{\rm cool,2}$ and $R_{\rm cool,2}\lesssim R\lesssim R_{\rm cool,1}$ is sketched in Figure \ref{fig:GRB}. For our fiducial parameters, we have $R\lesssim R_{\rm cool,2}$. Then the synchrotron spectrum has two spectral breaks at $E_{\rm b} = (2\theta_{\rm KN}/3\theta^3)E_{\rm s,pk}$ and $E_{\rm i} = (9/4)\theta^4E_{\rm s,pk}$,
\begin{align}
\label{eq:epsb}
E_{\rm b} & \sim 70\;L_{52}^{1/4} \Gamma_{300} R_{15}^{-1/2} E_6^{-1/2} \theta_{-1}^{-3/2}{\rm\; keV} \\
\label{eq:epsi}
E_{\rm i} & \sim 0.2\;\theta_{-1}^4 E_6{\rm\; keV} \;.
\end{align}
The spectral slopes are $\alpha=1$ for $E_{\rm i}\lesssim E\lesssim E_{\rm b}$, when cooling is dominated by IC in the Klein-Nishina regime, and $\alpha=1/2$ for $E_{\rm b}\lesssim E\lesssim E_{\rm s,pk}$, when cooling is dominated by synchrotron (note that $E_{\rm s,pk}/E_{\rm b}\sim 15$, and $E_{\rm b}/E_{\rm i}\sim 350$). If particles cool down to Lorentz factors $\gamma\lesssim (\theta_{\rm KN}/\theta)\sigma_{\rm e}$, a low energy break appears at $E_{\rm KN} = (\theta_{\rm KN}/\theta)^2 E_{\rm s,pk}$,
\begin{equation}
E_{\rm KN} \sim 1\; L_{52}^{1/2} \Gamma_{300}^2 R_{15}^{-1} E_6^{-2} \theta_{-1} {\rm\; eV}\;.
\end{equation}
The synchrotron spectral slopes are $\alpha=1/2$ for $E\lesssim E_{\rm KN}$, and $\alpha=3/4$ for $E_{\rm KN}\lesssim E\lesssim E_{\rm i}$. The spectral break at $E_{\rm KN}$ may be replaced by a cooling break if particles do not cool completely.

Large dissipation radii $R_{\rm cool,2}\lesssim R\lesssim R_{\rm cool,1}$ may be relevant for GRBs with large bulk Lorentz factors. For $\Gamma\sim 1000$, we find that $R_{\rm cool,1} \sim 10^{17}{\rm\; cm}$, and $R_{\rm cool,2} \sim 10^{16}{\rm\; cm}$. If $R_{\rm cool,2}\lesssim R\lesssim R_{\rm cool,1}$, the synchrotron spectrum has a cooling break at $E_{\rm cool} = (t_{\rm cool}^{\rm pk}/t_{\rm dyn})^2E_{\rm s,pk}$. Then
\begin{equation}
E_{\rm cool} \sim 50\; L_{52}^{-3/2}\Gamma_{1000}^6 R_{16} \theta_{-1}^{-3}{\rm\; keV} \;,
\end{equation}
where we have defined $\Gamma_{1000}\equiv\Gamma/1000$, and $R_{16}\equiv R/10^{16}{\rm\; cm}$.  Note that $E_{\rm cool}$ is much larger than in the isotropic case $\theta\sim 1$. The spectral slopes are $\alpha=1/2$ for $E_{\rm cool}\lesssim E\lesssim E_{\rm s,pk}$, and $\alpha=4/3$ (as usual for synchrotron radiation below the cooling break) for $E\lesssim E_{\rm cool}$. The soft part of the spectrum extends over a relatively narrow range of frequencies since $E_{\rm s,pk}/E_{\rm cool}\sim 20$ for the fiducial parameters of the model (the dependence of $E_{\rm s,pk}/E_{\rm cool}$ on the parameters is strong). Interestingly, many GRB spectra may be consistent with a broken power law with slopes $\alpha=4/3$ at low frequencies, and $\alpha=1/2$ close to the peak \citep[e.g.][]{Oganesyan+2017, Oganesyan+2018, Oganesyan+2019, Ravasio+2018, Ravasio+2019, Toffano+2021}.

We remark that synchrotron emission cannot produce very hard spectral slopes, $\alpha\gtrsim 4/3$.\footnote{The regime of extremely small pitch angles, $\theta\lesssim 1/\gamma$, is an exception to this general behaviour \citep[e.g.][]{LloydPetrosian2000, LloydPetrosian2002}. However, in this regime synchrotron radiation is extremely inefficient, making it difficult to produce the large luminosity of GRBs.} Fitting GRB spectra with empirical functions \citep[e.g.][]{Band1993} suggests that a significant fraction of GRBs have a low frequency slope $\alpha\gtrsim 4/3$, which violates the so-called synchrotron line-of-death \citep[e.g.][]{Preece+1998}. Another challenge for a synchrotron model is reproducing the sharpness of the Band function \citep[e.g.][]{AxelssonBorgonovo2015, Yu+2015}. However, these results have been recently questioned by fitting GRB spectra directly with synchrotron models \citep[e.g.][]{Burgess2019, Oganesyan+2019, Burgess+2020}.

\subsubsection{IC emission}

The total IC luminosity in the Klein-Nishina regime relevant for GRBs is a fraction $\eta=(2\theta_{\rm KN}/3\theta^3)^{1/2}$ of the synchrotron luminosity (see Table \ref{tab:sync3}). We have
\begin{equation}
\eta\sim 0.3\; L_{52}^{1/8} \Gamma_{300}^{1/2} R_{15}^{-1/4} E_6^{-3/4} \theta_{-1}^{-3/4} \;.
\end{equation}
If all the IC radiation escapes the system, the spectrum peaks at $E_{\rm IC, pk}=\Gamma\varepsilon_{\rm IC,pk}\simeq (1/2)\Gamma\sigma_{\rm e} m_{\rm e}c^2$, i.e.
\begin{equation}
E_{\rm IC, pk}\sim 4\; L_{52}^{-1/4}\Gamma_{300} R_{15}^{1/2} E_6^{1/2} \theta_{-1}^{-1/2} {\rm\; TeV} \;.
\end{equation}
The spectrum has two spectral breaks at $E_{\rm IC, b}= (2\theta_{\rm KN}/3\theta^3)^{1/2}E_{\rm IC,pk}$ and $E_{\rm IC, KN}= (\theta_{\rm KN}/\theta)E_{\rm IC,pk}$, i.e.
\begin{align}
E_{\rm IC, b} & \sim 1\; L_{52}^{-1/8}\Gamma_{300}^{3/2} R_{15}^{1/4} E_6^{-1/4} \theta_{-1}^{-5/4} {\rm\; TeV} \\
E_{\rm IC, KN} & \sim 4\; \Gamma_{300}^2 E_6^{-1} {\rm\; GeV} \;.
\end{align}
The spectral slopes are $\alpha=1/2$ for $E\lesssim E_{\rm IC,KN}$, $\alpha=1$ for $E_{\rm IC, KN}\lesssim E\lesssim E_{\rm IC, b}$, and $\alpha=0$ for $E_{\rm IC, b}\lesssim E\lesssim E_{\rm IC, pk}$. The spectral break at $E_{\rm IC, KN}$ may be replaced by a cooling break if the particles do not cool completely.

In the next section we show that IC photons with energy $E_{\rm IC}\gtrsim E_{\rm IC,KN}$ may easily annihilate and produce secondary pairs. Then only a small fraction $E_{\rm IC,KN}/E_{\rm IC,b}\sim 4\times 10^{-3}$ of the total IC luminosity escapes the system directly. Instead, most of the IC luminosity is transformed into kinetic energy of the secondary pairs.

\subsubsection{Pair production}
 \label{sec:pair}

The optical depth for pair production via photon-photon collisions is $\tau_{\gamma\gamma}=(\sigma_{\gamma\gamma}/\sigma_{\rm T})(8\theta\theta_{\rm KN}/3)^{1/2}\sigma_{\rm e} \ell_{\rm B}$ (see Table \ref{tab:sync3}). For a $\alpha=1$ spectrum of the target synchrotron photons, the cross section for photon-photon collisions is $\sigma_{\gamma\gamma}=(7/12)\sigma_{\rm T}$ \citep[e.g.][]{Svensson1987}. Then
\begin{equation}
\label{eq:tau}
\tau_{\gamma\gamma} \sim 60\; L_{52}^{7/8}\Gamma_{300}^{-5/2} R_{15}^{-3/4} E_6^{-1/4}\theta_{-1}^{3/4} \;.
\end{equation}
Pair production can be neglected if $\tau_{\gamma\gamma}\lesssim 1$, which gives $R\gtrsim R_{\gamma\gamma}$, where we have defined
\begin{equation}
\label{eq:Rdiss}
R_{\gamma\gamma}= 3\times 10^{17} L_{52}^{7/6}\Gamma_{300}^{-10/3} E_6^{-1/3}\theta_{-1} {\rm\; cm} \;.
\end{equation}
Note that $R_{\gamma\gamma}$ is a fraction $(\sigma_{\gamma\gamma}/\sigma_{\rm T})^{4/3}\sim 0.5$ of $R_{\rm cool,2}$ (compare Eqs. \ref{eq:Rdiss1} and \ref{eq:Rdiss}). At radii $R\lesssim R_{\gamma\gamma}$, the IC component should be efficiently reprocessed by the cascade of secondary electron-positron pairs, softening the spectrum of the IC component.

The secondary pairs also tend to soften the spectrum of the synchrotron component. This effect depends on the ratio $f_{\rm s}^{\rm sec}/f_{\rm s}^{\rm prim}$, where $f_{\rm s}=P_{\rm s}/(P_{\rm s}+P_{\rm IC})$  is the synchrotron fraction of the radiation emitted by the primary and secondary particles. If $f_{\rm s}^{\rm sec}\gtrsim f_{\rm s}^{\rm prim}$, the synchrotron spectrum emitted by the primary particles may be softened significantly. The ratio $f_{\rm s}^{\rm sec}/f_{\rm s}^{\rm prim}$ is controlled by the pitch angle of the secondary pairs ($f_{\rm s}^{\rm sec}$ may be larger than $f_{\rm s}^{\rm prim}$ if the secondary pairs have a pitch angle $\theta_{\rm sec}>\theta$).

Since IC photons annihilate after travelling a distance $l_\parallel=ct_{\rm dyn}/\tau_{\gamma\gamma}$ along the direction of the magnetic field, the pitch angle of the secondary pairs may be estimated as $\theta_{\rm sec}=\max[\theta, (\delta B/B)[l_\parallel]]$, where $(\delta B/B)[l_\parallel]$ is the amplitude of turbulent fluctuations at the scale $l_\parallel$. Assuming that the amplitude of turbulent fluctuations is $\delta B/B\propto l_\perp^{1/3}\propto l_\parallel^{1/2}$ \citep[e.g.][]{GoldreichSridhar1995, ThompsonBlaes1998}, we have $(\delta B/B)[l_\parallel]=s(l_\parallel/ct_{\rm dyn})^{1/2}=s\tau_{\gamma\gamma}^{-1/2}$ (the scaling constant $s$ is equal to the amplitude of the fluctuations at the scale of the largest turbulent eddy). Then $\theta_{\rm sec}=\max[\theta, s\tau_{\gamma\gamma}^{-1/2}]$. The secondary pairs have pitch angles comparable to the primary particles, i.e. $\theta_{\rm sec}\sim\theta$, if $s\tau_{\gamma\gamma}^{-1/2}\lesssim\theta$. For $\tau_{\gamma\gamma}\sim 60$ and $\theta\sim 0.1$, this condition is practically satisfied even for strong turbulent fluctuations, with $s\sim 1$. A lower level of fluctuations, $s<1$, is expected if turbulence develops from global instabilities of the jet \citep[e.g.][]{Davelaar+2020}.

\section{Conclusions}
\label{sec:conclusions}

In this paper we investigated the synchrotron-self-Compton radiation from magnetically-dominated turbulent plasmas in relativistic jets. Since observed relativistic jets have a high radiative efficiency, we considered fast cooling conditions, when particles radiate their energy on short timescales compared with the dynamical time of the jet expansion. Our model is motivated by recent first principles simulations of magnetically-dominated plasma turbulence, which show that electrons are impulsively accelerated to Lorentz factors $\gamma\sim\sigma_{\rm e}$ by reconnection in large-scale current sheets ($\sigma_{\rm e}$ is the plasma magnetisation, defined with respect to the electron rest mass energy density). Since the reconnection electric field is nearly aligned with the local magnetic field, the accelerated particles are strongly anisotropic.

The anisotropy has a strong impact on the spectrum of the emitted radiation. Since particles move nearly along the direction of the local magnetic field, synchrotron emission is suppressed. Then IC scattering may be the dominant cooling channel, even in magnetically-dominated plasmas. The synchrotron and IC spectra emitted by fast cooling particles are described by broken power laws (see Tables \ref{tab:sync1}-\ref{tab:sync6}). The slope of the power law segments is determined by the cooling regime (see Figure \ref{fig:param}). The most important features are summarised below.
\begin{itemize}
\item When the emitting electrons IC scatter the synchrotron radiation in the Thomson regime, the synchrotron and IC cooling times are inversely proportional to the particle Lorentz factor, i.e. $t_{\rm cool, s}\propto\gamma^{-1}$ and $t_{\rm cool, IC}\propto\gamma^{-1}$. The number of cooled particles per unit Lorentz factor is ${\rm d}n_{\rm e}/{\rm d}\gamma\propto\gamma^{-2}$, independent of the dominant cooling channel. Then synchrotron and IC radiation components have soft spectra, $\nu F_\nu\propto\nu^{1/2}$. In this regime, the ratio of the synchrotron and IC luminosities is $L_{\rm s}/L_{\rm IC}\sim\sin\theta\sim\theta$, where $\theta$ is the particle pitch angle (i.e. the angle between the particle velocity and the local magnetic field).
\item When the emitting electrons IC scatter the synchrotron radiation in the Klein-Nishina regime, the IC cooling time $t_{\rm cool, IC}$ typically approaches a constant independent of particle energy. For small particle Lorentz factors, IC is the dominant cooling channel. Then ${\rm d}n_{\rm e}/{\rm d}\gamma\propto t_{\rm cool, IC}/\gamma \propto\gamma^{-1}$, and synchrotron radiation has a hard spectrum, $\nu F_\nu\propto\nu$. For large particle Lorentz factors, IC cooling is strongly suppressed due to Klein-Nishina effects, and synchrotron becomes the dominant cooling channel. Then ${\rm d}n_{\rm e}/{\rm d}\gamma\propto t_{\rm cool, s}/\gamma \propto\gamma^{-2}$, and synchrotron radiation has a soft spectrum, $\nu F_\nu\propto\nu^{1/2}$.
\end{itemize}
We remark that the particle anisotropy is essential for the hardening of the synchrotron spectrum in magnetically-dominated plasmas. If particles are isotropic, synchrotron emission is inevitably the dominant cooling channel. Then both synchrotron and IC spectra are soft, $\nu F_\nu\propto\nu^{1/2}$.

We have applied our results to BL Lacs and GRB prompt emission, and found that synchrotron-self-Compton emission from anisotropic particles may be consistent with the observed spectra. Estimating the  required conditions inside the jet from the observed peak frequency and luminosity, we found that (i) the magnetic field strength in the plasma rest frame is $B\sim 1{\rm\; G}$ in BL Lacs, and $B\sim 1{\rm\; kG}$ in GRBs; (ii) electrons are accelerated to similar Lorentz factors, $\gamma\sim\sigma_{\rm e}\sim 10^4$. For electron-proton plasmas, $\sigma_{\rm e}\sim 10^4$ corresponds to an overall magnetisation $\sigma=(m_{\rm e}/m_{\rm p})\sigma_{\rm e}\sim 10$. 

In BL Lacs, electrons heated by magnetically-dominated turbulence IC scatter the synchrotron radiation in the Thomson regime. Then under fast cooling conditions synchrotron and IC components have soft spectra, $\nu F_\nu\propto\nu^{1/2}$. For pitch angles $\theta\gtrsim 0.1$, the synchrotron and IC luminosities are comparable (within a factor of ten), consistent with the properties of non-thermal radiation from BL Lacs. An exception to this general behaviour may be represented by orphan gamma-ray flares, i.e. IC flares with a negligible synchrotron counterpart. Since the ratio of the synchrotron and IC luminosities is $\sim\theta$, orphan gamma-ray flares may be produced when the particle distribution is extremely anisotropic (strongly anisotropic particles may produce orphan gamma-ray flares also in Flat Spectrum Radio Quasars; see \citealt[][]{Sobacchi+2021}). The pitch angle anisotropy may be regulated by (i) the level of the magnetic fluctuations from which turbulence develops. Larger fluctuations produce more isotropic particle distributions \citep[][]{Comisso+2020, Sobacchi+2021}; (ii) the plasma composition. In electron-proton plasmas, the anisotropy may be erased by kinetic instabilities that are absent in electron-positron plasmas \citep[][]{SobacchiLyubarsky2019}.

In GRBs, electrons heated by magnetically-dominated turbulence IC scatter the synchrotron radiation in the Klein-Nishina regime. For a peak frequency of the observed spectrum $h\nu_{\rm pk}\sim 1{\rm\; MeV}$, we find that IC is the dominant cooling channel for particles with a pitch angle $\theta\sim 0.1$ emitting at frequencies $0.2{\rm\; keV}\lesssim h\nu\lesssim 70{\rm\; keV}$. Then under fast cooling conditions the synchrotron radiation has a hard spectrum $\nu F_\nu\propto\nu$, consistent with a typical GRB. Synchrotron becomes the dominant cooling channel for particles emitting at frequencies $70{\rm\; keV} \lesssim h\nu\lesssim 1{\rm\; MeV}$. Then the synchrotron spectrum softens close to the spectral peak. The break frequency, $h\nu_{\rm b}\sim 70{\rm\; keV}$, moves close to the spectral peak when either $\nu_{\rm pk}$ or $\theta$ decrease (we find that $h\nu_{\rm b}\sim 130{\rm\; keV}$ for $h\nu_{\rm pk}\sim 300{\rm\; keV}$, and $h\nu_{\rm b}\sim 200{\rm\; keV}$ for $\theta\sim 0.05$).

There are aspects of our model that deserve further investigation. In GRBs, IC photons escaping from the emitting region may be observed at TeV energies. However, IC photons easily annihilate and produce electron-positron pairs. Although synchrotron radiation from the secondary pairs may be neglected under certain conditions (see Section \ref{sec:pair}), it is unclear whether these conditions occur in real GRB jets. We did not consider the reduction of the plasma magnetisation due to pair creation. A detailed study of this complicated issue is left for future work.

The peak energy and luminosity of the GRB prompt emission follow a well known correlation, $E_{\rm pk}\sim 0.3\;L_{\rm iso}^{1/2}{\rm\; MeV}$ \citep[e.g.][]{WeiGao2003, Yonetoku2004, Ghirlanda2012}. In our model, we find that $E_{\rm pk}\propto L_{\rm iso}^{1/2}\sigma_{\rm e}^2\theta/R$, where $R$ is the dissipation radius (see Eq. \ref{eq:sigmaeGRB}). Since variations of $\sigma_{\rm e}^2\theta/R$ tend to smear out the $E_{\rm pk}-L_{\rm iso}$ correlation, this quantity would need to be approximately constant among different bursts to reproduce a tight correlation. Similar issues regarding the origin of the $E_{\rm pk}-L_{\rm iso}$ correlation in magnetically-dominated GRB jets have been discussed by other authors \citep[e.g.][]{Lyutikov2006,ZhangYan2011}. On the other hand, the $E_{\rm pk}-L_{\rm iso}$ correlation may arise more naturally in photospheric emission models \citep[e.g.][]{Beloborodov2013}.

Our model describes the emitted spectrum only below the spectral peak, which is produced by particles injected with $\gamma\sim\sigma_{\rm e}$. Since we assumed that the acceleration timescale is a step function, $t_{\rm acc}\ll t_{\rm dyn}$ for $\gamma\sim\sigma_{\rm e}$ and $t_{\rm acc}\sim t_{\rm dyn}$ for $\gamma\gtrsim\sigma_{\rm e}$, fast cooling produces an exponential cutoff in the particle distribution for $\gamma\gtrsim\sigma_{\rm e}$. In a more realistic scenario, $t_{\rm acc}$ may have a smooth dependence on $\gamma$. Then particles can be accelerated up to a cutoff Lorentz factor $\gamma_{\rm co} \gtrsim\sigma_{\rm e}$, which is determined by the condition that the acceleration time is equal to the cooling time \citep[e.g.][]{NattilaBeloborodov2020}. Particles with $\sigma_{\rm e}\lesssim\gamma\lesssim \gamma_{\rm co}$ may be injected with a power law distribution ${\rm d}n_{\rm e}/{\rm d}\gamma\propto\gamma^{-p}$, with $p\sim 3$ \citep[e.g.][]{ComissoSironi2018, ComissoSironi2019}. This scenario may be consistent with the fact that BL Lac and GRB spectra are often described by a power law at frequencies larger than the peak frequency.

We assumed that the pitch angle is independent of the particle energy. This assumption is supported by first-principles simulations in fast cooling electron-positron plasmas \citep[][]{NattilaBeloborodov2020, Sobacchi+2021}. In electron-proton plasmas, pitch angle scattering due to kinetic instabilities may be more efficient for particles with small Lorentz factors, which has implications for the detailed modelling of BL Lac spectra \citep[][]{SobacchiLyubarsky2019, TavecchioSobacchi2020}. Simulations are needed to investigate the anisotropy of particles with Lorentz factors $\gamma \lesssim \sigma_{\rm e}$ in fast cooling electron-proton plasmas.

\section*{Acknowledgements}

We thank the anonymous referee for constructive comments and suggestions that improved the paper. We are grateful to Michael Burgess, Luca Comisso, Joonas N\"attil\"a and Fabrizio Tavecchio for insightful comments and discussions. LS acknowledges support from the Sloan Fellowship, the Cottrell Scholar Award, DoE DE-SC0016542,  NASA ATP 80NSSC18K1104, and NSF PHY-1903412. AMB acknowledges support from NSF grants AST 1816484 and AST 2009453, the Simons Foundation grant \#446228, and the Humboldt Foundation.

\section*{Data availability}

No new data were generated or analysed in support of this research.

\def\aap{A\&A}\def\aj{AJ}\def\apj{ApJ}\def\apjl{ApJ}\def\mnras{MNRAS}\def\prl{Phys. Rev. Lett.}
\def\araa{ARA\&A}\def\physrep{PhR}\def\sovast{Sov. Astron.}\def\nar{NewAR}\def\pasa{PASA}
\def\aapr{Astronomy \& Astrophysics Review}\def\apjs{ApJS}\def\nat{Nature}\def\na{New Astron.}
\def\prd{Phys. Rev. D}\def\pre{Phys. Rev. E}\def\pasp{PASP}\def\ssr{Space Sci. Rev.}
\bibliographystyle{mn2e}
\bibliography{2d}

\end{document}